\documentclass[%
 reprint,
 amsmath,amssymb,
 aps,
 pre,
]{revtex4-1}

\usepackage{graphicx}
\usepackage{dcolumn}
\usepackage{bm}
\usepackage{lipsum} 
\usepackage[colorlinks=true,linkcolor=blue]{hyperref}%
\usepackage{color}

\usepackage{CJK}
\usepackage{tabularx}
\usepackage{booktabs}
\usepackage{subfigure}
\usepackage{wrapfig}
\usepackage{verbatim}
\usepackage{mathrsfs}
\usepackage{multirow}

\begin{document}

\preprint{APS/123-QED}

\title{An inviscid model study of sandstorm in unstably stratified atmospheric boundary layer}
\author{Chenyue Xie}
\affiliation{CAPT-HEDPS, SKLTCS, Department of Mechanics and Engineering Science, College of Engineering, Peking University, Beijing 100871, People's Republic of China}%
\author{Xiangming Xiong}%
\affiliation{CAPT-HEDPS, SKLTCS, Department of Mechanics and Engineering Science, College of Engineering, Peking University, Beijing 100871, People's Republic of China}%
\author{Jianjun Tao}
\email{Email: jjtao@pku.edu.cn}
\affiliation{CAPT-HEDPS, SKLTCS, Department of Mechanics and Engineering Science, College of Engineering, Peking University, Beijing 100871, People's Republic of China}%

\date{\today}

\begin{abstract}
According to field observations, the atmospheric boundary layer  is usually unstably stratified before a dust and sandstorm, the particle-laden turbulent gravity current with an extremely high Reynolds number. In this paper, an inviscid model is built to study the mechanism governing the slumping phase of gravity current, and it is shown that the dimensionless current front speed, the Froude number, decreases when the current fluid or the ambient medium or both fluids are unstably stratified. In spite of  the density interface mixing, the relation between the front speed and the front height described by the inviscid model agrees with the numerical simulation results, where the lock-exchange gravity currents with different initial lock heights are calculated for different unstable stratification cases. Furthermore, the velocity increments obtained by field observations at the sandstorm fronts are satisfactorily consistent with the evaluations of the model, suggesting that the inviscid mechanism makes contribution to such high Reynolds number turbulent flows.
\end{abstract}

\pacs{Valid PACS appear here}
\maketitle

\section{\label{sec:level1}INTRODUCTION}
Gravity currents are flows of one fluid into an ambient fluid caused by density differences, e.g. sandstorm and  avalanche, and are importance in geophysics, atmospheric physics, and environmental fluid mechanics  \cite{Benjamin1968,Simpson1999,Meiburg2010}. The propagation of gravity current undergoes several stages: the acceleration stage, the slumping stage, and the self-similar deceleration stage. A key characteristic of the current is its front speed, which  is almost constant at the slumping stage and was studied in the non-Boussinesq currents  \cite{Lowe2005}, the partial-depth gravity currents  \cite{Shin2004}, and the intrusive gravity currents  \cite{Maurer2010}. By assuming that both fluids are homogeneous and inviscid, a steady theory was developed first by Benjamin  \cite{Benjamin1968} for the front speed and the current height, and was extended for stably stratified ambient fluids with linear \cite{Ungarish2006} and nonlinear \cite{White2008} density profiles, where the front speeds were found to be smaller than their counterparts in homogeneous ambient fluids.  In order to consider the effects of shear and lateral heat of condensation, a nonlinear model was proposed for the gravity current \cite{Liu1996}, where the temperature difference between the current and the ambient fluid was assumed to be proportional to the ambient temperature and the ambient temperature was  assumed to be an exponential increasing function of the vertical position.  The dissipationless conjugate-state solutions of the steady theory \cite{White2008} were found to be consistent with the experimental results of boundary gravity currents  \cite{Maxworthy2002} and intrusive gravity currents \cite{Britter1981,Faust1984}. An alternative vorticity-based steady model was proposed  \cite{Borden2013} and extended to consider Boussinesq gravity currents in sheared and stably stratified ambient fluids \cite{Azadani2015, Azadani2016} and non-Boussinesq gravity currents \cite{Konopliv2016}. Homogeneous and stably stratified gravity currents and intrusive gravity currents in stably stratified ambient fluids were studied as well with one-layer shallow-water models \cite{Ungarish2002, Ungarish2005a, Ungarish2005b, Ungarish2012, Goldman2014}. Recently, a Rayleigh-Taylor model was proposed for the gravity currents  to explain the formation mechanism of lobe and cleft structures at the sandstorm front \cite{Xie2019,Zhang2021}. In contrast to the various models for currents within homogeneous or stably stratified fluids, the corresponding study for unstable stratification is still rudimental.

Dust and sandstorms are strong gravity currents and devastating natural hazards causing desertification and pollution \cite{Global2001}. According to field observations, the local wind velocity increases gradually and the atmospheric boundary layer  is unstably stratified before the particle concentration in the wind increases substantially \cite{Liu2021}. Because the thermal capacities of sand and earth are generally smaller than that of the air, the ground temperature increases faster than the air temperature does under sunshine, and the atmosphere may illustrate inverse or unstable stratification in sunny days.  The buoyancy effect becomes important for unstably stratified atmosphere boundary layers, and its influence on the mean velocity and temperature profiles are studied by the Monin-Obukhov similarity theory \cite{Monin1954}.  It was shown that within an unstably stratified boundary layer there are three sublayers, where the temperature obeys power laws \cite{Kader1990}. Recent numerical simulations and field observations of unstably stratified atmospheric boundary layers revealed that the mean potential temperature remains a logarithmic function of the vertical position in a wide range of the Monin-Obukhov stability parameter \cite{Cheng2021}. Unstably stratified atmosphere without wind corresponds to turbulent Rayleigh-B\'{e}nard convection, which also has a logarithmic temperature profile \cite{Ahlers2014,He2021}. By now,  sandstorms are still difficult to be simulated completely with laboratory experiments and direct numerical simulations due to the high Reynolds number, multi-phase manner, and multi-field properties. Therefore, it is necessary to develop dynamic models, which  adapt conveniently for different sublayers and density stratification, to investigate the dominant characteristics of sandstorms.

\section{PHYSICAL MODEL AND INVISCID THEORY}
\begin{figure}[h]
\centering
\includegraphics[scale=0.78]{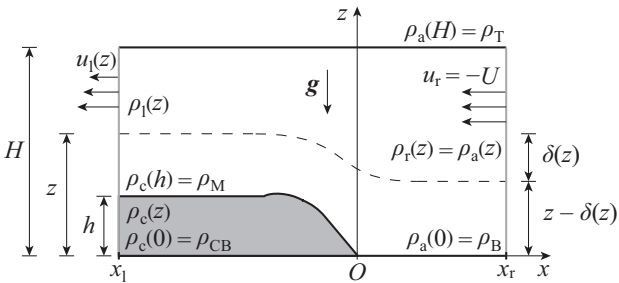}
\caption{Schematic plot of the gravity current in the reference frame moving with the current front velocity. The shaded region represents the current and the dashed line is a representative streamline.}
\label{Fig1}
\end{figure}

A schematic plot of the gravity current model considered in this paper is shown in Fig.\,\ref{Fig1}, and the whole flow field is steady because the coordinate system moving at the current front speed $U$ in the $x$ direction. The density profiles in the far upstream ambient fluid and the far downstream current are functions of the vertical coordinate $z$,  $\rho_\mathrm{a}(z)$ and $\rho_\mathrm{c}(z)$, respectively. Note that the downstream side is in the $-x$ direction.

When the density difference between the two fluids is small, the two-dimensional Navier-Stokes equations with the Boussinesq approximation are
\begin{align}
& \frac{\partial u}{\partial x} + \frac{\partial w}{\partial z} = 0, \label{eq:continuity} \\
& \frac{\partial u}{\partial t} + u \frac{\partial u}{\partial x} + w \frac{\partial u}{\partial z} = -\frac{1}{\rho_0} \frac{\partial p}{\partial x} + \nu \nabla^2 u, \label{eq:momentum_x} \\
& \frac{\partial w}{\partial t} + u \frac{\partial w}{\partial x} + w \frac{\partial w}{\partial z} = -\frac{1}{\rho_0} \frac{\partial p}{\partial z} - \frac{\rho}{\rho_0} g + \nu \nabla^2 w, \label{eq:momentum_z} \\
& \frac{\partial \rho}{\partial t} + u \frac{\partial \rho}{\partial x} + w \frac{\partial \rho}{\partial z} = \kappa \nabla^2 \rho, \label{eq:density}
\end{align}
where $u$, $w$, $\rho$, and $p$ are the horizontal velocity, vertical velocity, fluid density, and pressure, respectively. $\rho_0$ is a reference density and will be prescribed later. $g$,  $\nu$, and $\kappa$  are the magnitude of the gravitational acceleration, the kinematic viscosity, and the density diffusion coefficient of the fluids, respectively.

The minimum density in the unstably stratified ambient fluid is $\rho_\mathrm{B} = \rho_\mathrm{a}(0)$, which is the density of the far upstream ambient fluid at the bottom boundary. The maximum density in the unstably stratified current is $\rho_\mathrm{M} = \rho_\mathrm{c}(h)$, where $h$ is the height of the gravity current. The stratification parameters  of the ambient fluid and the current are defined as
\begin{align}
& S = \frac{\rho_\mathrm{T} - \rho_\mathrm{B}}{\rho_\mathrm{M} - \rho_\mathrm{B}} = \frac{\rho_\mathrm{a}(H) - \rho_\mathrm{a}(0)}{\rho_\mathrm{c}(h) - \rho_\mathrm{a}(0)}, \label{eq:S} \\
& \sigma = \frac{\rho_\mathrm{M} - \rho_\mathrm{CB}}{\rho_\mathrm{M} - \rho_\mathrm{B}} = \frac{\rho_\mathrm{c}(h) - \rho_\mathrm{c}(0)}{\rho_\mathrm{c}(h) - \rho_\mathrm{a}(0)}, \label{eq:sigma}
\end{align}
respectively, where $\rho_\mathrm{T} = \rho_\mathrm{a}(H)$ is the density of the far upstream ambient fluid at the top boundary and $\rho_\mathrm{CB} = \rho_\mathrm{c}(0)$ is the density of the far downstream current at the bottom boundary. For the present model we have $\rho_\mathrm{M} \geq \rho_\mathrm{T}$ and $\rho_\mathrm{CB} \geq \rho_\mathrm{B}$, and hence the maximum density difference of the flow system is  $\rho_\mathrm{M} - \rho_\mathrm{B}$. The unstably stratified ambient fluid and current correspond to $0 < S < 1$ and $0 < \sigma < 1$, respectively.

The Froude number is defined as
\begin{equation}
Fr = \frac{U}{\sqrt{\hat{g} h}}, \label{eq:Fr}
\end{equation}
where the reduced gravity $\hat{g}$ is defined as
\begin{align}
& \hat{g} = \epsilon g, \label{eq:g'} \\
& \epsilon = \frac{\rho_\mathrm{M} - \rho_\mathrm{B}}{\rho_\mathrm{0}} \ll 1. \label{eq:epsilon}
\end{align}

As an inviscid model, the momentum and density diffusion effects are ignored as in the previous studies \cite{Benjamin1968,Ungarish2006, White2008}. Consequently, Eqs.\,\eqref{eq:momentum_x}--\eqref{eq:density} are simplified as
\begin{align}
& u \frac{\partial u}{\partial x} + w \frac{\partial u}{\partial z} = -\frac{1}{\rho_0} \frac{\partial p}{\partial x}, \label{eq:momentum_x_conservation} \\
& u \frac{\partial w}{\partial x} + w \frac{\partial w}{\partial z} = -\frac{1}{\rho_0} \frac{\partial p}{\partial z} - \frac{\rho}{\rho_0} g, \label{eq:momentum_z_conservation} \\
& u \frac{\partial \rho}{\partial x} + w \frac{\partial \rho}{\partial z} = 0. \label{eq:density_conservation}
\end{align}

At the bottom and top walls, no-penetration boundary conditions are applied:
\begin{equation}
w|_{z=0} = w|_{z=H} = 0, \label{eq:bc_no_penetration}
\end{equation}
and at the far downstream and upstream positions, $x_\mathrm{l}$ and $x_\mathrm{r}$, the following boundary conditions are used as shown in Fig.\,\ref{Fig1}:
\begin{align}
& \rho_\mathrm{l}(z) \equiv \rho(x_\mathrm{l}, z) = \rho_\mathrm{c}(z), \quad \ \ 0 \leq z < h, \label{eq:rho_l_c} \\
& u_\mathrm{l}(z) \equiv u(x_\mathrm{l}, z) = 0, \qquad \quad \; 0 \leq z < h, \label{eq:u_l_c} \\
& w(x_\mathrm{l}, z) = \frac{\partial w}{\partial x}(x_\mathrm{l}, z) = 0, \quad \, 0 \leq z \leq H, \label{eq:w_l} \\
& \rho_\mathrm{r}(z) \equiv \rho(x_\mathrm{r}, z) = \rho_\mathrm{a}(z), \quad \ \, 0 \leq z \leq H, \label{eq:rho_r} \\
& u_\mathrm{r}(z) \equiv u(x_\mathrm{r}, z) = -U, \qquad \, 0 \leq z \leq H, \label{eq:u_r} \\
& w(x_\mathrm{r}, z) = \frac{\partial w}{\partial x}(x_\mathrm{r}, z) = 0, \quad  0 \leq z \leq H. \label{eq:w_r}
\end{align}
The density and velocity in the far downstream ambient fluid are derived in the Appendix as follows,
\begin{align}
& \rho_\mathrm{l}(z) \equiv \rho(x_\mathrm{l}, z) = \rho_\mathrm{a}(z - \delta(z)), \qquad h < z \leq H, \label{eq:rho_l_a} \\
& u_\mathrm{l}(z) \equiv u(x_\mathrm{l}, z) = -U[1 - \delta'(z)], \quad \, h < z \leq H, \label{eq:u_l_a}
\end{align}
where $'$ represents $d/dz$ and $\delta(z)$ is the streamline displacement (Fig.\,\ref{Fig1}) and satisfies
\begin{align}
& \delta''(z) = \frac{g}{\rho_0 U^2} \frac{\mathrm{d} \rho_\mathrm{a}}{\mathrm{d} z}|_{z - \delta(z)} \delta(z), \qquad h < z \leq H, \label{eq:DJL} \\
& \delta(h) = h, \label{eq:DJL_bc1} \\
& \delta(H) = 0. \label{eq:DJL_bc2}
\end{align}

Next, we derive the relationship between the front speed $U$ and the current height $h$ from the momentum equations. Integrating  Eq.\,\eqref{eq:momentum_x_conservation} and Eq.\,\eqref{eq:momentum_z_conservation} and using the boundary conditions, we have
\begin{align}
 &p_\mathrm{r}(z) \equiv p(x_\mathrm{r}, z) = p_\mathrm{r}(0) - g \int_0^z \rho_\mathrm{a}(z') \, \mathrm{d}z', \label{eq:p_r}\\
 &p(x_\mathrm{l}, 0) = p(x_\mathrm{r}, 0) + \frac{1}{2} \rho_0 U^2, \label{eq:p_l_p_r}\\
 &p_\mathrm{l}(z) \equiv p(x_\mathrm{l}, z) = p_\mathrm{r}(0) + \frac{1}{2} \rho_0 U^2 - g \int_0^z \rho_\mathrm{l}(z') \, \mathrm{d}z', \label{eq:p_l}
\end{align}
when $0 \leq z \leq H$.

Integrating Eq.\,\eqref{eq:momentum_x_conservation} in the region $[x_\mathrm{l}, x_\mathrm{r}] \times [0, H]$ and using Eqs.\,\eqref{eq:continuity} and \eqref{eq:bc_no_penetration}, we have
\begin{equation}
\int_0^H (p_\mathrm{l} + \rho_0 u_\mathrm{l}^2) \, \mathrm{d}z = \int_0^H (p_\mathrm{r} + \rho_0 u_\mathrm{r}^2) \, \mathrm{d}z. \label{eq:momentum_x_integration}
\end{equation}

Substituting Eqs.\,\eqref{eq:rho_l_c}--\eqref{eq:u_l_a}, \eqref{eq:p_r}, and \eqref{eq:p_l} into Eq.\,\eqref{eq:momentum_x_integration} and using Eqs.\,\eqref{eq:DJL}--\eqref{eq:DJL_bc2}, we have
\begin{align}
& \frac{1}{4} \rho_0 U^2 \int_h^H \delta'^3 \, \mathrm{d}z - g \int_0^h (H - z) \rho_\mathrm{c}(z) \, \mathrm{d}z + \rho_\mathrm{B} g h (H - \frac{h}{2}) \nonumber \\
& + \frac{1}{2} \rho_0 U^2 (H - \frac{h}{2}) [1 - \delta'(h)]^2 + \frac{1}{4} \rho_0 U^2 h=0, \label{eq:flow_force_balance}
\end{align}
where $\rho_\mathrm{a}(0) = \rho_\mathrm{B}$ is used. Eq.\,\eqref{eq:flow_force_balance} coupled with   Eqs.\,\eqref{eq:DJL}--\eqref{eq:DJL_bc2} describes the relation between the front speed $U$ and the current height $h$ for arbitrary and independent, stable and unstable stratifications of  $\rho_a$ and $\rho_c$.   When  $\rho_\mathrm{c}$ is a constant (uniform current),  $\rho_{0}=\rho_{B}$, and $\rho_{B}>\rho_{T}$  (stable stratification), Eq. \eqref{eq:flow_force_balance} is equivalent to Eq.\,(4.8) of Reference  \cite{White2008}. 

\section{GRAVITY CURRENT OF LINEARLY STRATIFIED FLUIDS}

In this section, the model is simplified for linear density distributions as
\begin{align}
& \rho_\mathrm{c}(z) = \rho_\mathrm{CB} + (\rho_\mathrm{M} - \rho_\mathrm{CB}) \frac{z}{h}, \quad \, 0 \leq z \leq h, \label{eq:rho_c_linear} \\
& \rho_\mathrm{a}(z) = \rho_\mathrm{B} + (\rho_\mathrm{T} - \rho_\mathrm{B}) \frac{z}{H}, \qquad 0 \leq z \leq H, \label{eq:rho_a_linear}
\end{align}
and we set  $\rho_0 = \rho_\mathrm{B}$. Substituting Eq.\,\eqref{eq:rho_a_linear} into Eq.\,\eqref{eq:DJL}, we have
\begin{equation}
\delta''(z) = \frac{\beta^2}{H^2} \delta(z), \qquad h < z \leq H, \label{eq:DJL_linear}
\end{equation}
where
\begin{equation}
\beta = \frac{1}{U} \sqrt{\frac{|\rho_\mathrm{T} - \rho_\mathrm{B}| g H}{\rho_\mathrm{B}}}.  \label{eq:beta} \\
\end{equation}

\subsection{Unstably stratified ambient fluid}
In an unstably stratified ambient fluid with $0 < S \leq 1$ and $\rho_\mathrm{T} > \rho_\mathrm{B}$, the solution of Eq.\,\eqref{eq:DJL_linear} under the boundary conditions \eqref{eq:DJL_bc1} and \eqref{eq:DJL_bc2} is
\begin{equation}
\delta(z) = h \frac{\sinh[\beta (1 - z/H)]}{\sinh[\beta (1 - h/H)]}, \qquad h \leq z \leq H. \label{eq:delta}
\end{equation}
It is noted that for stably stratified fluids, $\rho_\mathrm{T} < \rho_\mathrm{B}$, $-\beta^2$ instead of $\beta^2$ should be used in Eq.\,\eqref{eq:DJL_linear}, and $\delta(z)$ is then a trigonometric function as in the previous study \cite{Ungarish2006}.

Substituting Eqs.\,\eqref{eq:rho_c_linear} and \eqref{eq:delta} into Eq.\,\eqref{eq:flow_force_balance} and using Eqs.\,\eqref{eq:S}--\eqref{eq:epsilon} and \eqref{eq:beta}, we have
\begin{align}
& a^2 \gamma^2 \coth^2 \gamma + a(2-a) \gamma \coth \gamma + 1 - a \nonumber \\
& - \frac{\gamma^2}{S} \frac{a}{1-a} [2 - a - S\frac{a^2}{3} - \sigma (1 - \frac{a}{3})] = 0, \label{eq:flow_force_balance_linear}
\end{align}
where
\begin{equation}
 a = \frac{h}{H}, \ \  \gamma = \beta (1-a) . \label{eq:a} \\
\end{equation}

For given non-dimensional current height $a$ and stratification parameters $S$ and $\sigma$, $\gamma$ can be solved from Eq.\,\eqref{eq:flow_force_balance_linear} and $U$ is obtained from  Eq.\,\eqref{eq:beta}. Consequently, the Froude number is 
\begin{equation}
Fr(a, S, \sigma) = \frac{1-a}{\gamma} \sqrt{\frac{S}{a}}. \label{eq:Fr_stratified}
\end{equation}

\subsection{Homogeneous ambient fluid}
For a homogeneous or an unstably stratified gravity current ($0 \le \sigma \leq 1$) in a homogeneous ambient fluid ($S = 0$), we have $\beta = 0$ according to Eq.\,\eqref{eq:beta}. The solution of Eq.\,\eqref{eq:DJL_linear} under the boundary conditions \eqref{eq:DJL_bc1} and \eqref{eq:DJL_bc2} is
\begin{equation}
\delta(z) = h \frac{H-z}{H-h}, \qquad h \leq z \leq H. \label{eq:delta_homogeneous}
\end{equation}

Substituting Eq.\,\eqref{eq:rho_c_linear}, \eqref{eq:delta_homogeneous} and $\rho_\mathrm{a} = \rho_\mathrm{B}$ into Eq.\,\eqref{eq:flow_force_balance} and using Eqs.\,\eqref{eq:sigma}--\eqref{eq:epsilon}, we have
\begin{equation}
Fr(a,0,\sigma) = \sqrt{\frac{1-a}{1+a} [2 - a - \sigma (1 - \frac{a}{3})]}. \label{eq:Fr_homogeneous}
\end{equation}

If the current is also homogeneous ($\sigma = 0$), then it follows from Eq.\,\eqref{eq:Fr_homogeneous} that
\begin{equation}
Fr(a,0,0) = \sqrt{\frac{(1-a)(2-a)}{1+a}}, \label{eq:Fr_Benjamin}
\end{equation}
and the Benjamin theory is recovered  \cite{Benjamin1968}.

\section{NUMERICAL SIMULATIONS}

\begin{figure}[h]
\centering
\includegraphics[scale=0.75]{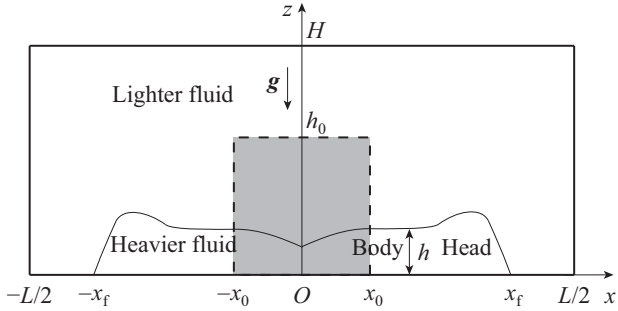}
\caption{Schematic plot of the lock-exchange gravity current between free-slip walls. The shaded region is initially occupied by the denser fluid. }
\label{Fig2}
\end{figure}

In order to validate the inviscid model, two-dimensional numerical simulations of the lock-exchange gravity current are carried out as shown in Fig.\,\ref{Fig2}. The denser fluid occupies initially a rectangular region with a length of $2x_0$ and a height of $h_0 \leq H$, where $H$ is the distance between the free-slip walls. The length and height of the computational domain are $L$ and $H$, respectively.

\begin{figure*}[ht]
	\centering \mbox{
		\subfigure[$S = 0$,   $\sigma = 0$]{\includegraphics[scale=0.41]{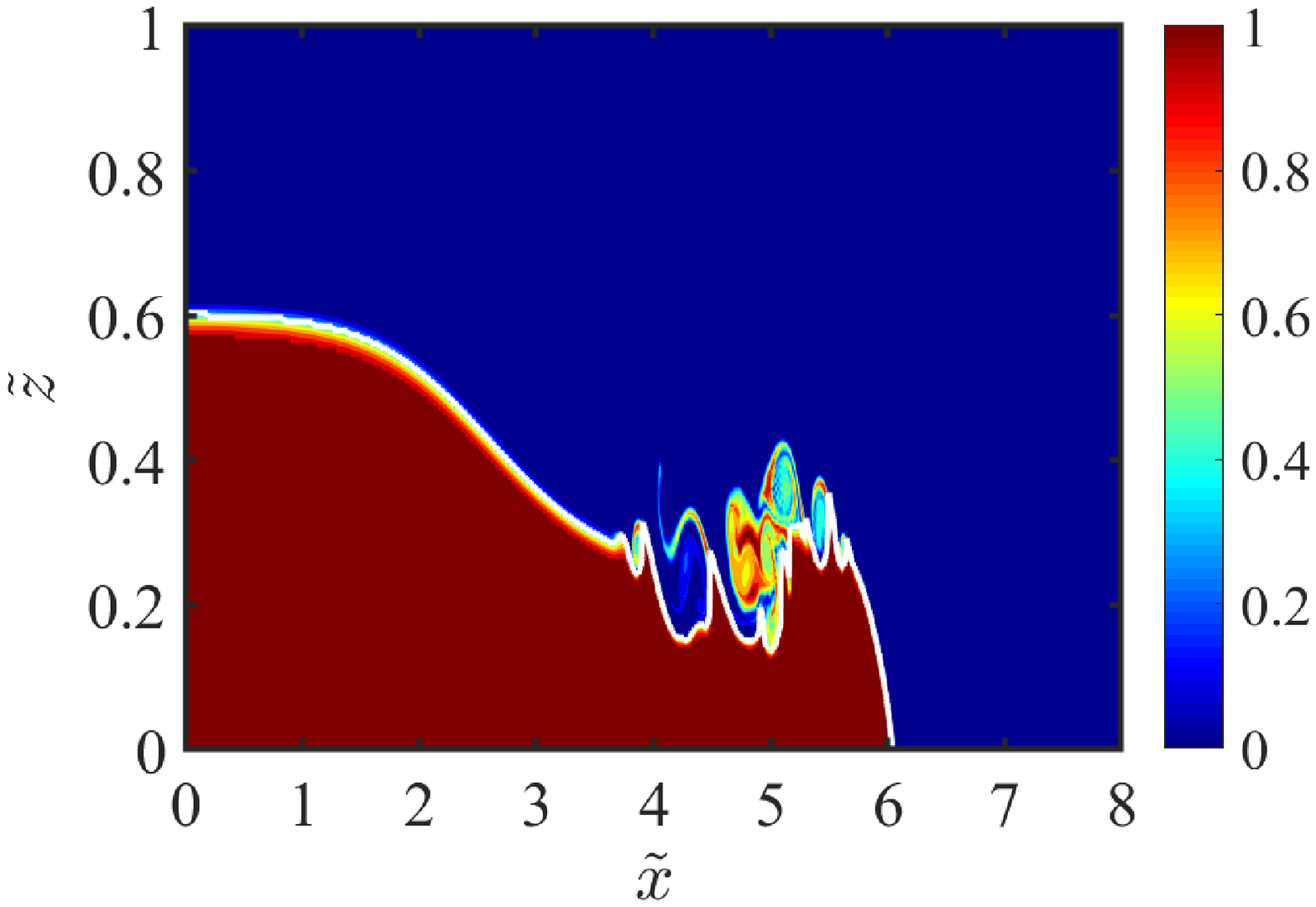}} \quad
		\subfigure[$S = 0.5$, $\sigma = 0.2$]{\includegraphics[scale=0.41]{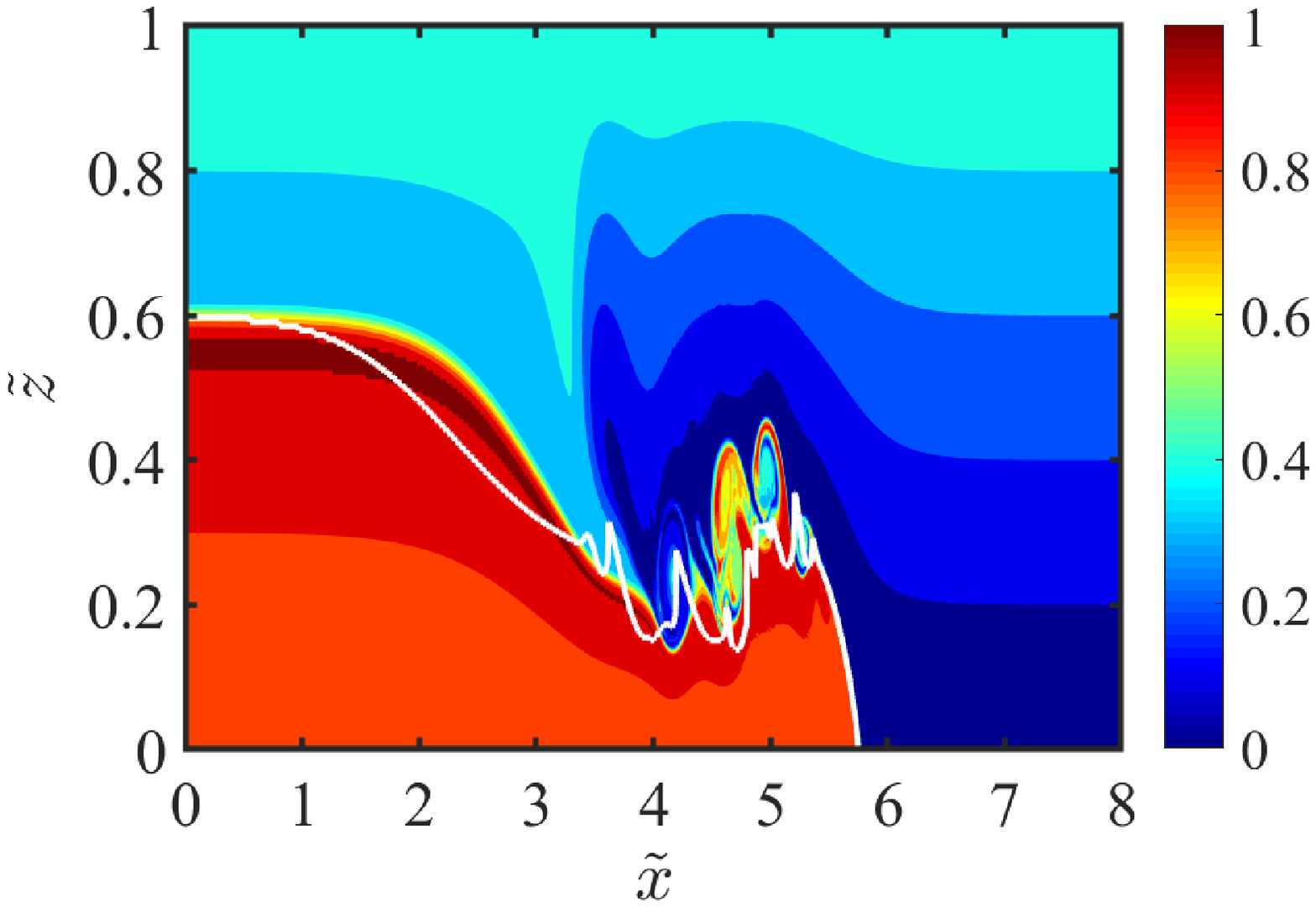}}}
	\centering \mbox{
		\subfigure[$S = 0.5$, $\sigma = 0.5$]{\includegraphics[scale=0.41]{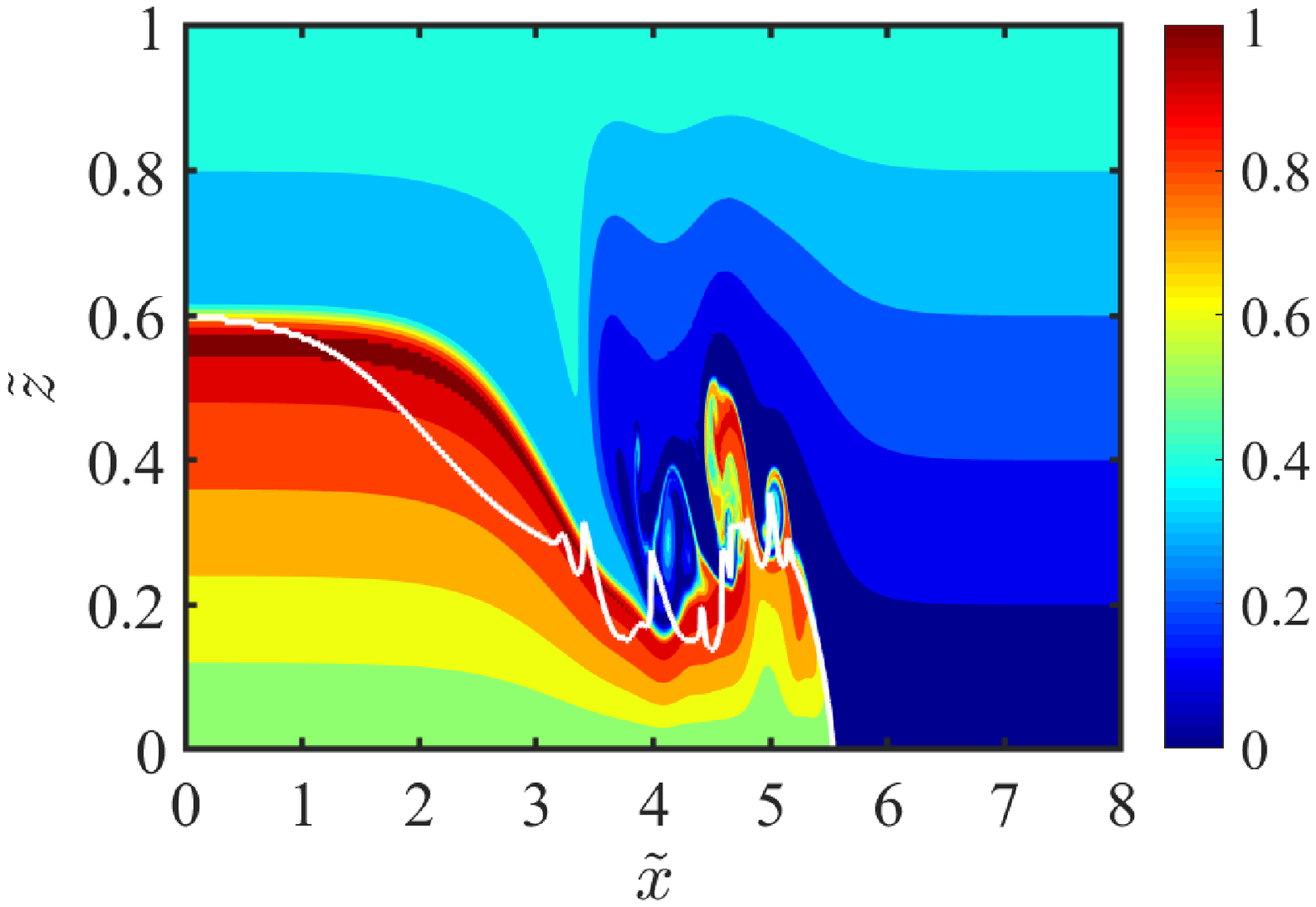}} \quad
		\subfigure[$S = 0.5$, $\sigma = 0.8$]{\includegraphics[scale=0.41]{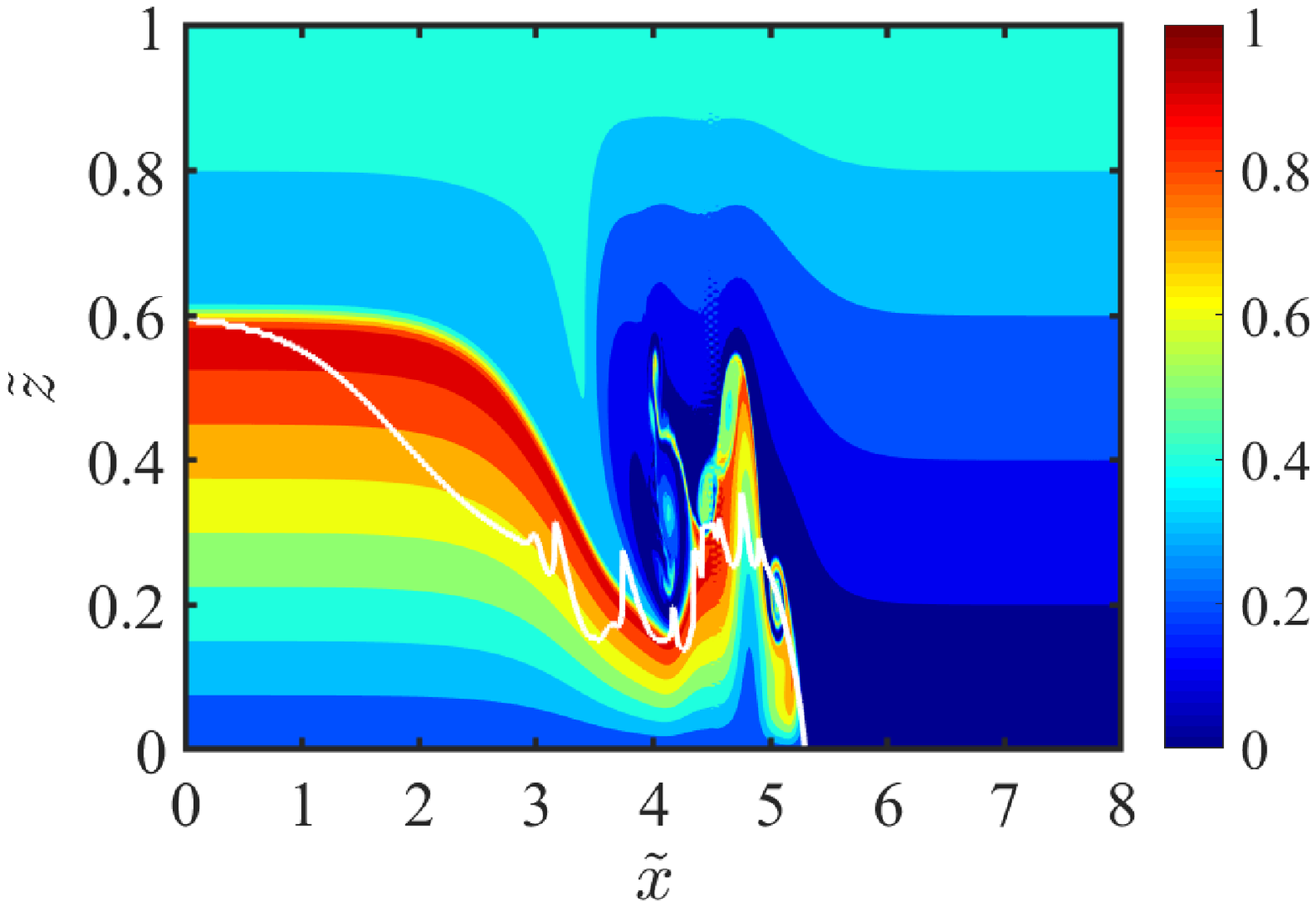}}}
	\caption{Contours of the density $\tilde{\rho}$ at $\tilde{t} = 10^{-4}$ in the numerical simulations with $Gr = 2 \times 10^9$ and $\tilde{h}_0 = 0.6$. The solid white curve shown in (a) indicates the local current height $\tilde{h}$, and is shifted accordingly in (b), (c), and (d) to fit the most downstream point of the current head.} \label{Fig3}
\end{figure*}

\begin{figure}
	\centering
	\includegraphics[scale=0.3]{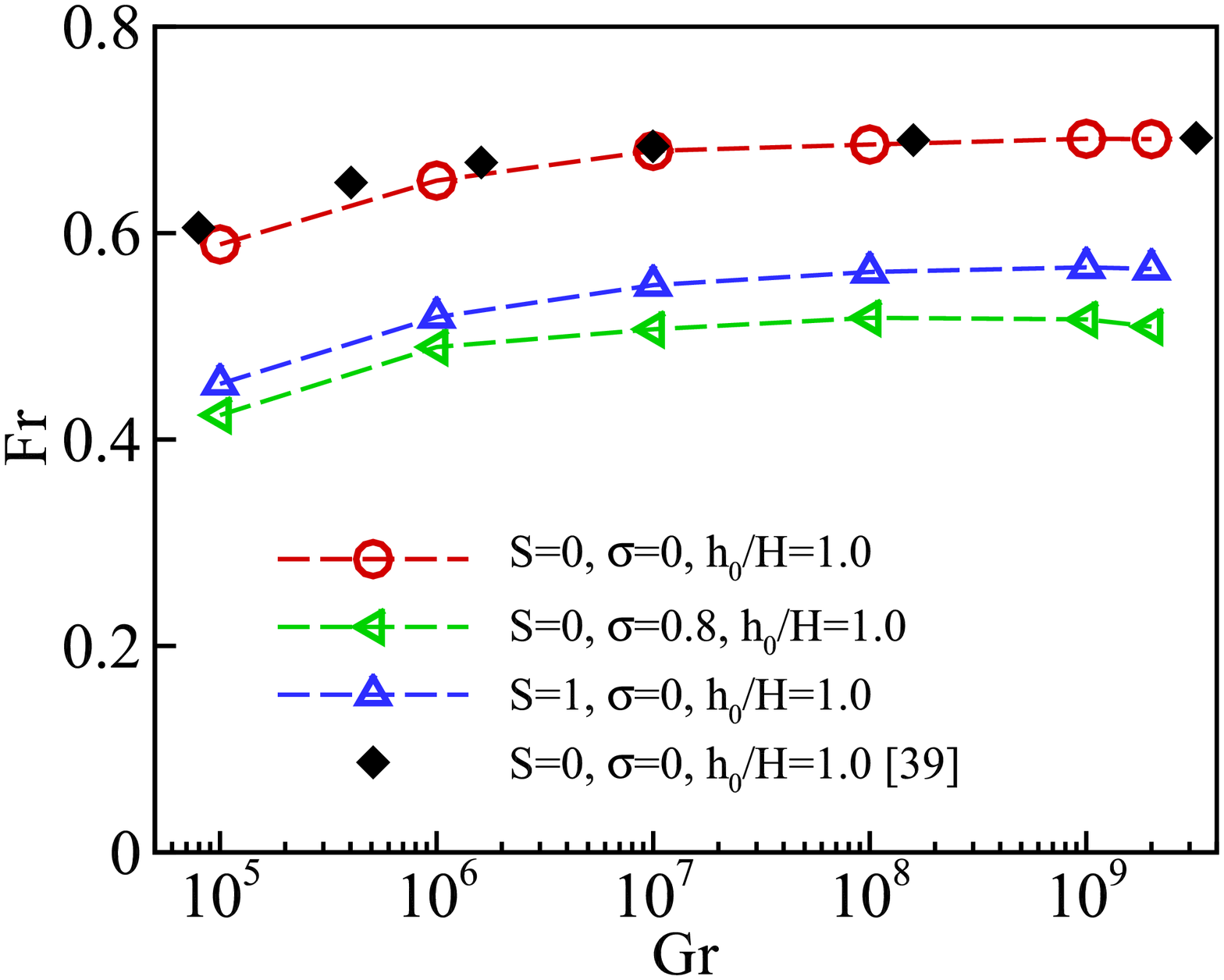}
	\caption{The Froude number $Fr$ as a function of the Grashof number $Gr$ in the numerical simulations. The hollow symbols represent the present numerical results, and the numerical results of \cite{Hartel2000} are shown as filled diamonds for reference. }
	\label{Fig4}
\end{figure}

\begin{figure*}[ht]
	\centering \mbox{
		\subfigure[$\sigma = 0$]{\includegraphics[scale=0.37]{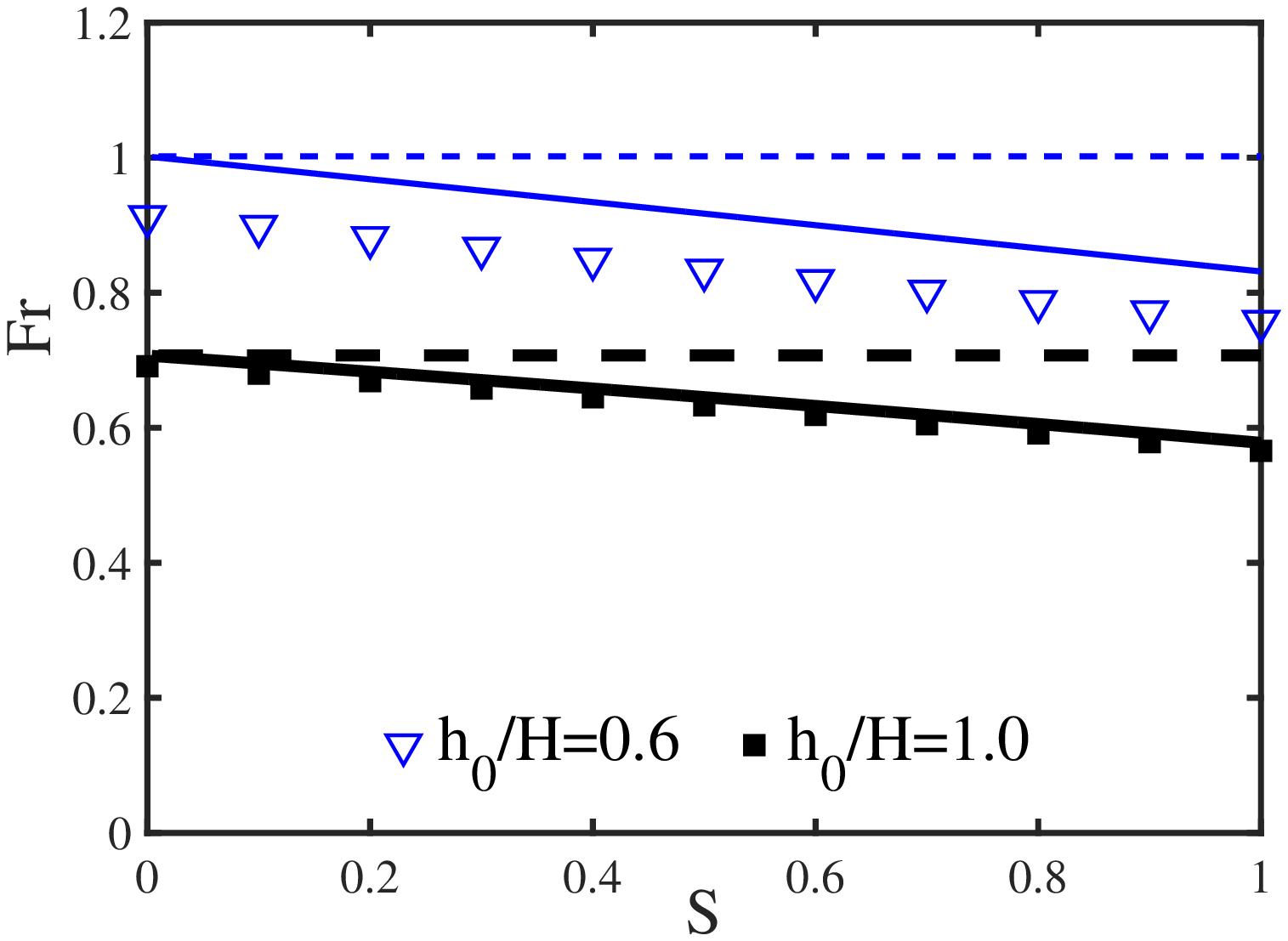}} \quad
		\subfigure[$\sigma = 0.2$]{\includegraphics[scale=0.37]{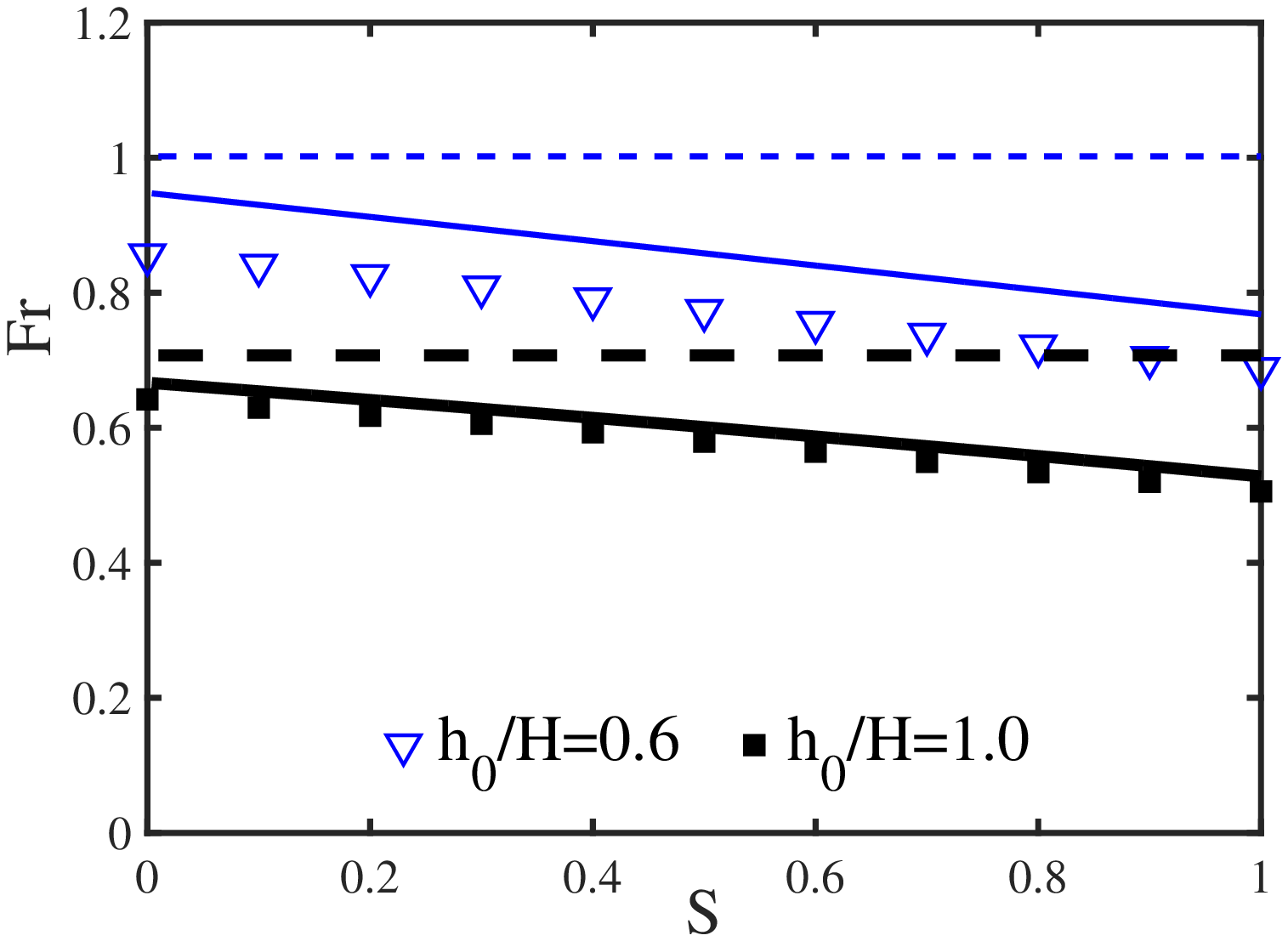}}}
	\centering \mbox{
		\subfigure[$\sigma = 0.5$]{\includegraphics[scale=0.37]{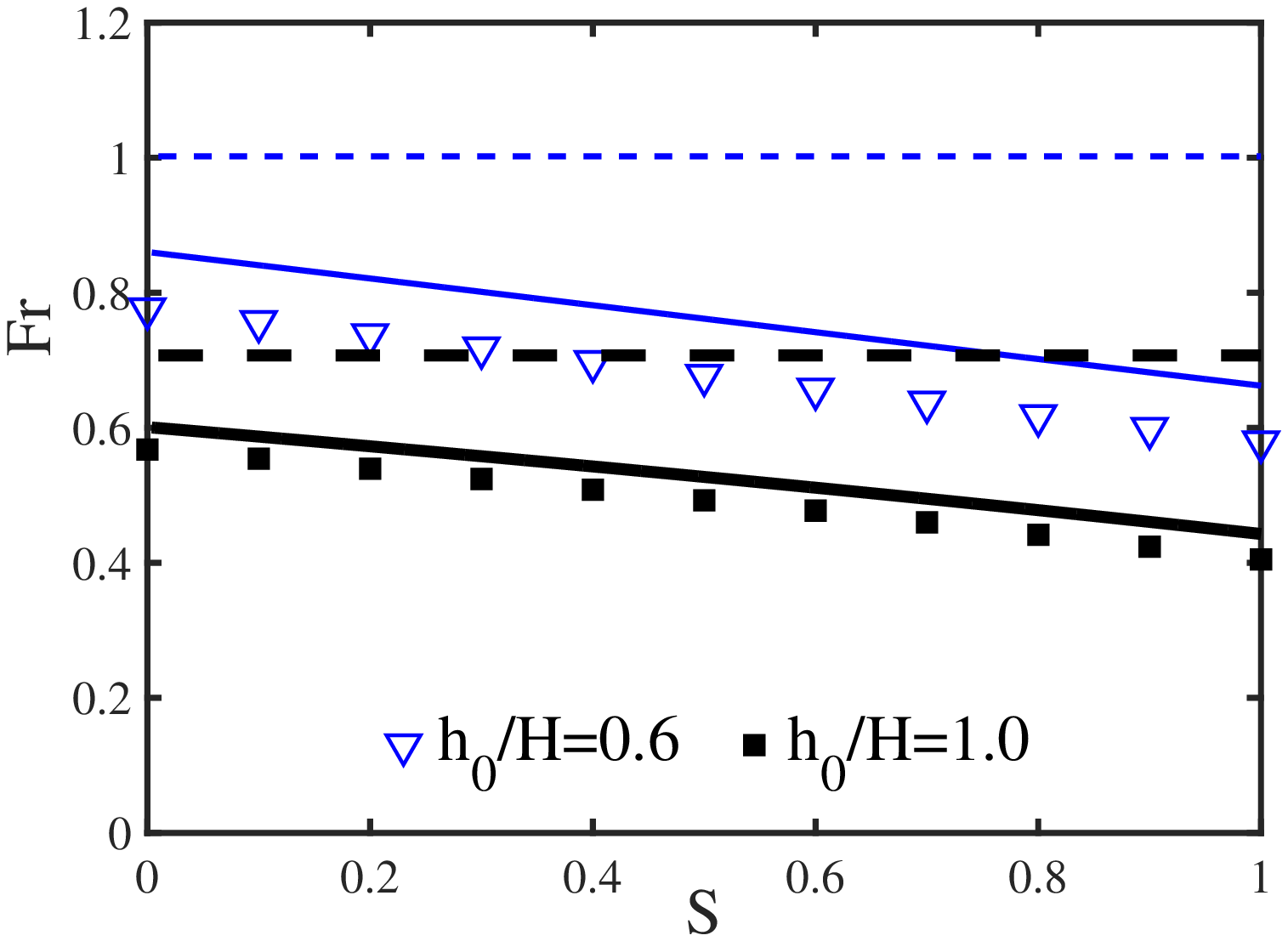}} \quad
		\subfigure[$\sigma = 0.8$]{\includegraphics[scale=0.37]{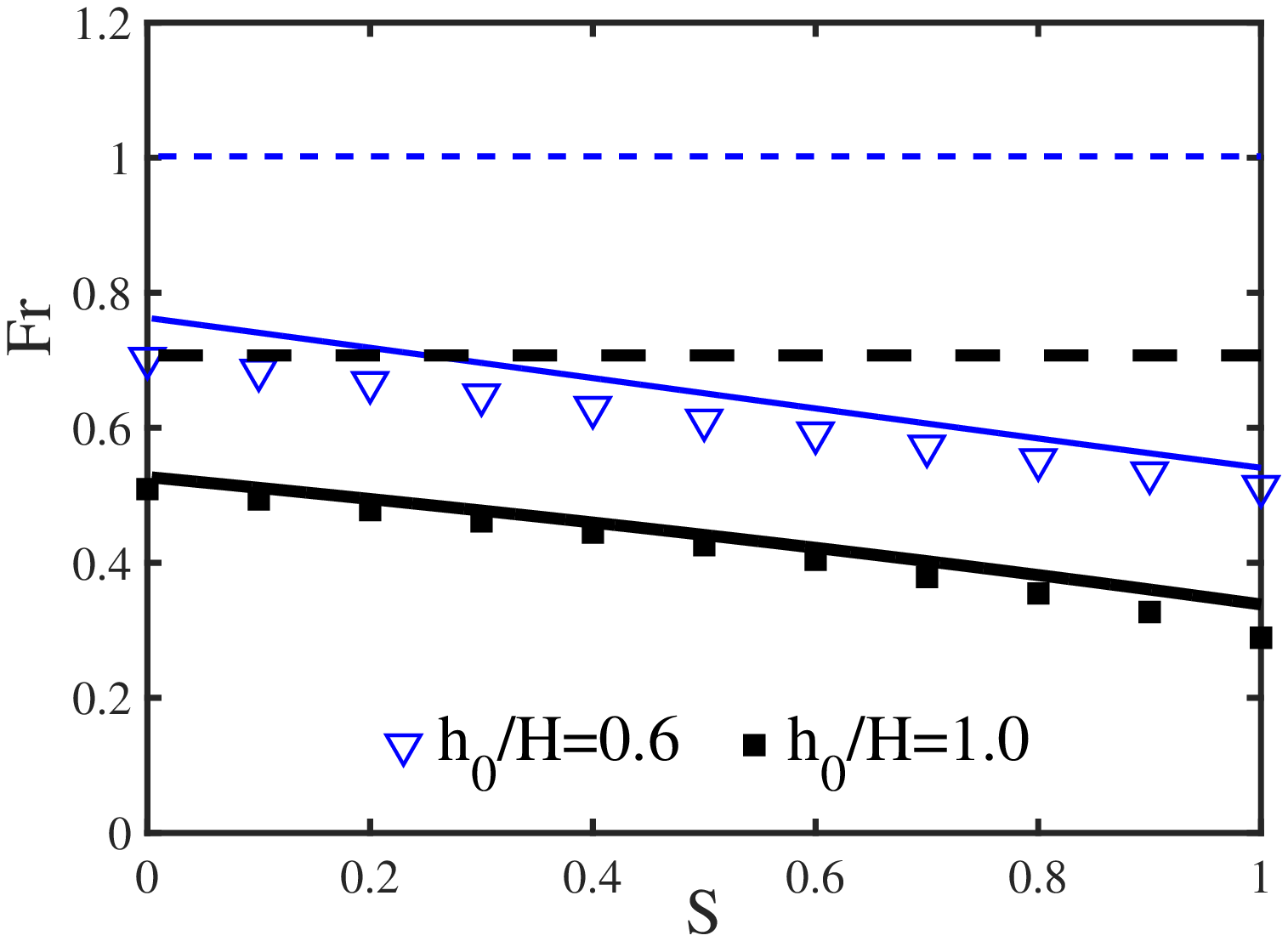}}}
	\caption{The Froude number $Fr$ as a function of $S$, $\sigma$, and $\tilde{h}_0=h_0/H$ at $Gr = 2 \times 10^9$. The symbols represent the numerical results. The solid and dashed lines represent the results of the present theory (Eq.\,\eqref{eq:Fr_stratified}) and the Benjamin's theory \cite{Benjamin1968} (Eq.\,\eqref{eq:Fr_Benjamin}), respectively.} \label{Fig5}
\end{figure*}

Introducing the non-dimensional coordinates, time, velocity, density, and pressure
\begin{align}
& (\tilde{x}, \tilde{z}) = \frac{(x, z)}{H}, \quad \tilde{t} = \frac{\nu}{H^2} t, \quad (\tilde{u}, \tilde{w}) = \frac{H}{\nu} (u, w), \quad \nonumber \\
& \tilde{\rho} = \frac{\rho - \rho_0}{\rho_\mathrm{M} - \rho_0}, \quad \tilde{p} = \frac{H^2}{\rho_0 \nu^2} (p + \rho_0 g z)
\end{align}
into Eqs.\,\eqref{eq:continuity}--\eqref{eq:density}, we have the non-dimensional governing equations:
\begin{align}
& \frac{\partial \tilde{u}}{\partial \tilde{x}} + \frac{\partial \tilde{w}}{\partial \tilde{z}} = 0, \label{eq:continuity_dimensionless} \\
& \frac{\partial \tilde{u}}{\partial \tilde{t}} + \tilde{u} \frac{\partial \tilde{u}}{\partial \tilde{x}} + \tilde{w} \frac{\partial \tilde{u}}{\partial \tilde{z}} = -\frac{\partial \tilde{p}}{\partial \tilde{x}} + \frac{\partial^2 \tilde{u}}{\partial \tilde{x}^2} + \frac{\partial^2 \tilde{u}}{\partial \tilde{z}^2}, \\
& \frac{\partial \tilde{w}}{\partial \tilde{t}} + \tilde{u} \frac{\partial \tilde{w}}{\partial \tilde{x}} + \tilde{w} \frac{\partial \tilde{w}}{\partial \tilde{z}} = -\frac{\partial \tilde{p}}{\partial \tilde{z}} - Gr \tilde{\rho} + \frac{\partial^2 \tilde{w}}{\partial \tilde{x}^2} + \frac{\partial^2 \tilde{w}}{\partial \tilde{z}^2}, \\
& \frac{\partial \tilde{\rho}}{\partial \tilde{t}} + \tilde{u} \frac{\partial \tilde{\rho}}{\partial \tilde{x}} + \tilde{w} \frac{\partial \tilde{\rho}}{\partial \tilde{z}} = \frac{1}{Sc} (\frac{\partial^2 \tilde{\rho}}{\partial \tilde{x}^2} + \frac{\partial^2 \tilde{\rho}}{\partial \tilde{z}^2})  \label{eq:density_dimensionless}
\end{align}
with
\begin{equation}
Gr = \frac{(\rho_\mathrm{M} - \rho_0) g H^3}{\rho_0 \nu^2} = \frac{\hat{g} H^3}{\nu^2}, \label{eq:Gr}
\end{equation}
where we have used $\rho_0 = \rho_\mathrm{B}$ and Eqs.\,\eqref{eq:g'} and \eqref{eq:epsilon}. The Schmidt number is $Sc = \nu / \kappa$ and is assumed to be 1 in the present simulations.

Equations \eqref{eq:continuity_dimensionless}--\eqref{eq:density_dimensionless} are solved with a pseudo-spectral method \cite{Chevalier2007} under the periodic boundary conditions in the horizontal direction. In the vertical direction, we assume the free-slip boundary conditions for the velocity and the zero-flux boundary condition for the density. The governing equations are discretized with 4096 Fourier modes and 257 Chebyshev polynomials in the horizontal and vertical directions, respectively. To achieve a high volume of release with $x_0 \gg h$ \cite{Constantinescu2014}, we choose $\tilde{x}_0 = x_0 / H = 4$ and $\tilde{L} = L / H = 16$. Note that the mode resolutions used in the $x$ and $z$ directions are 4 times and 1.25 times as large as those used in the previous study  \cite{Bonometti2008}, respectively, and it is checked that the front velocities obtained with half modes agreed with the present results to better than $3\%$.

Both fluids are at rest initially. The initial density distribution is
\begin{align}
\tilde{\rho} (\tilde{x}, \tilde{z}, 0) = \left \{\aligned
& 1 - \sigma + \sigma \tilde{z} / \tilde{h}_0, \quad |\tilde{x}| \leq \tilde{x}_0, \ 0 \leq \tilde{z} \leq \tilde{h}_0, \\
& S \tilde{z}, \qquad \qquad \qquad \mathrm{elsewhere},
\endaligned
\right. \label{eq:rho_initial}
\end{align}
where $\tilde{h}_0 = h_0 / H$. The non-dimensional numerical front speed is defined as $\tilde{U} = UH / \nu = \mathrm{d} \tilde{x}_\mathrm{f} / \mathrm{d} \tilde{t}$, where $\tilde{x}_\mathrm{f}$ is the abscissa of the intersection of the bottom wall and the isopycnal $\tilde{\rho} = (1 - \sigma)/2$ \cite{Birman2007}. The Froude number is therefore $Fr = \tilde{U} / \sqrt{a Gr}$ according to Eqs.\,\eqref{eq:Fr}, \eqref{eq:a}, and \eqref{eq:Gr}.

Based on the energy conservation for a steady current of inviscid homogeneous fluids ($S = \sigma = 0$), Benjamin proposed that the dimensionless current height is $a=1/2$  \cite{Benjamin1968}. Due to the velocity difference between the denser and lighter fluids near the interface, Kelvin-Helmholtz type instability may occur and the interface is mostly unstable. In addition, viscous dissipation will affect the current height as well. Consequently, different definitions were applied for the current height in the literature, e.g. the maximum height of the current head  \cite{Simpson1979} and a temporal and spatial average over some distance behind the current head  \cite{Benjamin1968,Borden2012,Borden2013}. It has been shown that $Fr$ and $h$ predicted by the steady and inviscid Benjamin theory are consistent generally with experiments  \cite{Huppert1980,Hacker1996} and numerical simulations \cite{Hartel2000} for currents of homogeneous fluids. Considering the presence of interface mixing, these consistencies are striking  \cite{Hacker1996} and may be explained  as follows. Firstly, though the instabilities and mixing occur near the density interface following a current front, the far upstream is undisturbed and at the downstream position where $\tilde{h}=\tilde{h}_0/2$, e.g. $\tilde{x} \approx 3.3$ in Fig. 3(a), the interface is not mixed. Consequently, the upstream and downstream boundary conditions still can be approximated by the inviscid model (Eqs.\,\eqref{eq:rho_l_c}--\eqref{eq:w_r}) during the slumping phase. Secondly, according to the experiments, the current front remains essentially unmixed during the slumping phase, and following the front there is a tail region near the bottom, which is almost unaffected by the mixing around the upper interface  \cite{Hallworth1993,Hallworth1996}. Similar phenomena can be observed in Fig. 3 as well. Thirdly, the characteristic current velocity ($\sqrt{\hat{g}h}$) is much faster than the diffusion ones (e.g. $\nu/h$) for a current with small $\nu$, and the error caused by ignoring the diffusion effects is limited during the slumping phase. Therefore, the inviscid theory does represent the basic mechanism governing the slumping phase, where the front velocity and the current height are assumed as constants.

Inspired by these previous studies, the non-dimensional numerical current height $a$ for $S = \sigma = 0$  is defined as,
\begin{align}
& a = \frac{1}{\tilde{t}_2-\tilde{t}_1}\int_{\tilde{t}_1}^{\tilde{t}_2} \frac{1}{\tilde{x}_d-\tilde{x}_u} \int_{\tilde{x}_u}^{\tilde{x}_d}\tilde{h}\mathrm{d} \tilde{x}\mathrm{d} \tilde{t},\ \ \ \nonumber \\
& \tilde{h}(\tilde{x}, \tilde{t})= \int_0^1 \tilde{\rho}(\tilde{x}, \tilde{z}, \tilde{t}) \, \mathrm{d} \tilde{z},
\end{align}
where $\tilde{h}$ is the local current height, $\tilde{x}_u(\tilde{t})$ and $\tilde{x}_d(\tilde{t})$ are the most upstream and downstream positions where $\tilde{h}=\tilde{h}_0/2$, respectively, and $(\tilde{t}_1, \tilde{t}_2)$ is a period in the slumping phase. According to the above definition of $a$, Benjamin's solution $a=1/2$ is recovered when $h_0=H$ due to the symmetric property of the interface. The contours of the density $\tilde{\rho}$  for different stratification cases at the same $\tilde{t}$, $Gr$, and $\tilde{h}_0$ are shown in Fig.\,\ref{Fig3}, and the front speed is substantially affected by $S$ and $\sigma$: increasing the unstably stratification parameter of the current $\sigma$ decreases the front speed.

According to the simulations shown in Fig. 3, the current interfaces for different stratified cases look similar as the solid white curve, the $\tilde{h}$ curve for  homogeneous case (Fig. 3a) shifted to fit the front heads. Consequently, the entrainment characteristics of the present stratified cases are similar as those for homogeneous fluids. Because of the interface mixing and diffusion, $\tilde{h}$ varies with time and the streamwise location, but its spatially averaged value does not change markedly during the slumping phase. Therefore, the current height $a$ for a stratified case is set approximately the same value as that of a homogeneous case with the same $Gr$ and $h_0$.

In order to verify the present inviscid model,  the viscosity should be small or $Gr$ should be large enough in the numerical simulations. However, large $Gr$ for unstably stratified fluids will lead to Rayleigh-B\'{e}nard type convection and disrupt the horizontal current.  Before the  Rayleigh-B\'{e}nard type convection takes part in and plays a serious role,  the currents simulated in this paper experience  steady slumping phases, which  are fortunately long enough to verify the inviscid model.
As shown in Fig.\,\ref{Fig4}, $Fr$ increases with $Gr$ at moderate Grashof numbers and remains nearly constant as $Gr>10^9$. Therefore, $Gr = 2 \times 10^9$ is used in the following simulations, and the corresponding non-dimensional current heights for homogeneous fluids are $a = 0.27$ and $0.50$ when the initial lock heights are $\tilde{h}_0 = 0.6$ and $1$, respectively.

According to the simulations, $a$ is less than $\tilde{h}_0/2$ when $\tilde{h}_0<1$, and the corresponding $Fr$ obtained with the Benjamin's theory (Eq.\,\eqref{eq:Fr_Benjamin}) increases with the decrease of $\tilde{h}_0$, illustrating the same trend as the simulations for the homogeneous fluids ($\sigma = S = 0$) as shown in Fig.\,\ref{Fig5}(a). For the unstably stratified fluids, however, the Benjamin's theory is not applicable, because $Fr$ becomes a decreasing function of $S$ and $\sigma$. It is shown in Fig.\,\ref{Fig5} that for different $S$ and $\sigma$, the present theoretical predictions shown by the solid lines agree with the simulation results, which are shown as symbols in Fig.\,\ref{Fig5}. The stratification effects may be explained as follows. According to Eq.\,\eqref{eq:rho_initial}, the average density of the ambient fluid increases with $S$, while the average density of the current decreases with the increase of $\sigma$. Consequently, the increases of $S$ and $\sigma$ decrease the average density difference between the two fluids, the source of driving force for gravity currents, and hence reduce the front speed and $Fr$. 

\section{SANDSTORM}

The Reynolds numbers of  sandstorms are extremely high ($Re_{\tau}\sim 10^6$ or higher),  and are at least two orders of magnitude  higher than  the present ability of direct numerical simulations  \cite{Cheng2021}.  Considering that the current velocities ($Fr$) and heights predicted by the steady and inviscid Benjamin theory are consistent generally with those of turbulent gravity currents obtained in experiments  \cite{Huppert1980, Hacker1996}, we use the present inviscid model to estimate the characteristic velocities of sandstorms.

It is found in field observations that before the sandstorm, the wind velocity is small (less than 3 m/s) and the Monin-Obukhov stability parameter may be less than  $-0.5$, indicating an unstable stratification  \cite{Liu2021}. The wind becomes stronger with time, and then the sand concentration experiences a remarkable growth, indicating the front of the sandstorm. As shown in Fig. 2 of \cite{Li2021}, when the PM10 (particles with size less than 10 $\mu$m) concentration measured at a height of 3.49m increased markedly from about zero (at  time of 11:00) to a value around 3.8 mg/m$^3$ (at 12:30), the wind streamwise velocity was enhanced simultaneously from  6 m/s in average to a saturated value of 7.7 m/s at a temperature around 16$^o$C, indicating that the sandstorm reached its long-term slumping phase (lasting about 4 hours),  where the present modal may be applicable. Since the streamwise velocity fluctuation and the particle concentration have strong  positive correlation in the vertical direction \cite{Wang2017}, the velocity increment 1.7 m/s is related to the density increase and corresponds to the current velocity $U$ discussed in the model. It was found that the mean density in a sandstorm was nearly constant at heights between 10 m and 30 m \cite{Zhang2021}.  By assuming that the PM10 percentage at the measurement site is about $1\%$, the  sand concentration near the front of  sandstorm \cite{Li2021} is estimated as 380  mg/m$^3$.

It is difficult to measure the heights of sandstorms. According to the brightness temperature of
the thick peak on satellite images, it was reported that the height of a strong sandstorm occurred in northern China in 1993 was about 2,200 meters \cite{Global2001}. Based on micro-pulse lidar observations, it was revealed that a dust layer in south Asia peaked at a height about 1300 m and extended up to about 3000 m above ground level \cite{Srivastava2011}. Considering that fine sand dust of 0.05-0.005 mm can be blown up to 1500 m \cite{Global2001}, the height of sandstorm \cite{Li2021} may be evaluated as 1300m.  In the inviscid model, the Froude number is a function of the density stratifications and $h/H$ or $h_0/H$.  As shown in Fig. 5 (a) and 5(b), the gravity currents with parameters between $(h_0/H, \sigma, S)=(0.6, 0, 0)$ and $(1, 0.2, 0.1)$ correspond to a Froude number in the range of $0.65 \sim 1$. By substituting the evaluated current height 1300 m,  the density difference 380 mg/m$^3$, and the air density at  16$^o$C into Eq.\,\eqref{eq:Fr}, it is easy to find that $Fr=0.65\sim 1$ corresponds to a current front velocity of $U=1.30\sim 1.99$ m/s, which is  consistent acceptably with 1.7 m/s, the  velocity increment contributed by the density difference of the sandstorm as discussed above.

Similarly, we may analyze the data of another sandstorm \cite{Liu2021}. As the PM10 concentration measured at a height of 5 m increased remarkably from 0 at 15:05 to 4.3 mg/m$^3$ at 16:40, the corresponding wind streamwise velocity increased from 6.3 m/s to 8.6 m/s. The velocity increment due to density difference is about 2.3 m/s and the sand concentration of the sandstorm is estimated as 430 mg/m$^3$. By using the air density at  13.4$^o$C and the same estimated current height 1300 m,  it is found that $Fr=0.65\sim 1$ corresponds to a current front velocity between $1.37\sim 2.11$ m/s, agreeing generally with the field observation value 2.3 m/s.  It is noted that the present inviscid model is built for strong sandstorms with sharp front interfaces (sand walls), while the sand concentration of  the two cases discussed above varies intermittently with time  at the front regions, a phenomenon correlated with the very large scale motion or gusty wind  \cite{Wang2017}. 

\section{CONCLUSIONS}
Sandstorms are  multi-phase wall-bounded turbulent flows with extremely high Reynolds numbers, and hence are difficult to be studied by direct numerical simulations and laboratory experiments. In order to understand the prominent features of sandstorms, an inviscid model is developed especially for the unstable stratification, which is usually the precursor of sandstorms.  The relation between the front velocity and the current height is derived for fluids with  linearly unstable stratification, and is confirmed by the numerical simulations of lock-exchange gravity currents, indicating that the unstable stratification retards the current velocity and decreases the Froude number. By applying the model to sandstorms, it is shown that the estimated velocity increments at the fronts of sandstorms agree acceptably with the field measurements. Considering the strong turbulent diffusion and dissipation in the atmospheric boundary layers, the general consistency between the field observation and the inviscid model seems surprising at the first glance, but reveals the basic inviscid mechanism: the velocity increment at the front of sandstorm is intrinsically related to the driving source, the density difference.  In reality, many factors will influence the kinematic and dynamic properties of sandstorms, such as  meteorological conditions, surface plants and buildings, and topographic features. In order to improve the model and our understanding of these extreme weather events, more field observations are expected and new technologies should be developed to measure the sandstorm height and multi-field properties in the future.

\begin{acknowledgments}
The simulations were performed on TianHe-1(A), and the support from the National Natural Science Foundation of China is acknowledged (Grants No. 91752203 and 11490553).
\end{acknowledgments}

\section{APPENDIX STREAMLINE DISPLACEMENT EQUATION}

According to Eq.\,\eqref{eq:density_conservation}, the density is constant along each streamline, and therefore $\rho = \rho(\psi)$, where $\psi$ is the stream function such that
\begin{align}
& u = -\frac{\partial \psi}{\partial z}, \label{eq:stream_function_u}\\
& w = \frac{\partial \psi}{\partial x}.
\end{align}
Without loss of generality, we require $\psi(x_\mathrm{r}, 0) = 0$. Then we have
\begin{equation}
\psi(x_\mathrm{r}, z) = Uz, \qquad 0 \leq z \leq H \label{eq:psi_r}
\end{equation}
according to Eqs.\,\eqref{eq:u_r} and \eqref{eq:stream_function_u}.

Eliminating the pressure in Eqs.\,\eqref{eq:momentum_x_conservation} and \eqref{eq:momentum_z_conservation} and using Eqs.\,\eqref{eq:continuity} and \eqref{eq:density_conservation}, we have
\begin{equation}
u \frac{\partial f}{\partial x} + w \frac{\partial f}{\partial z} = 0,
\end{equation}
where
\begin{equation}
f(x,z) = \frac{\partial u}{\partial z} - \frac{\partial w}{\partial x} - \frac{gz}{\rho_0} \frac{\mathrm{d} \rho}{\mathrm{d} \psi}.
\end{equation}
The function $f(x,z)$ is therefore constant along each streamline.

For any point $(x_\mathrm{l}, z)$ in the ambient fluid ($h \leq z \leq H$) at the left boundary in Fig.\,\ref{Fig1}, we define a streamline displacement $\delta(z)$ such that the points $(x_\mathrm{l}, z)$ and $(x_\mathrm{r}, z - \delta(z))$ are on the same streamline. Consequently, we have $\rho(x_\mathrm{l}, z) = \rho(x_\mathrm{r}, z - \delta(z))$, and then
\begin{equation}
\rho_\mathrm{l}(z) \equiv \rho(x_\mathrm{l}, z) = \rho_\mathrm{a}(z - \delta(z)), \qquad h < z \leq H \label{eq:rho_l_a_appendix}
\end{equation}
according to Eq.\,\eqref{eq:rho_r}.

Similarly, we have $f(x_\mathrm{l}, z) = f(x_\mathrm{r}, z - \delta(z))$, and then it follows from Eqs.\,\eqref{eq:w_l}, \eqref{eq:u_r}, and \eqref{eq:w_r} that
\begin{align}
& \frac{\mathrm{d} u_\mathrm{l}}{\mathrm{d} z}  = \frac{gz}{\rho_0} \frac{\mathrm{d} \rho}{\mathrm{d} \psi}(x_\mathrm{l},z) - \frac{g[z - \delta(z)]}{\rho_0} \frac{\mathrm{d} \rho}{\mathrm{d} \psi}(x_\mathrm{r}, z - \delta(z)), \nonumber \\
& h < z \leq H. \label{eq:f_conservation}
\end{align}

As the stream function $\psi$ is constant along each streamline, we have $\psi(x_\mathrm{l}, z) = \psi(x_\mathrm{r}, z - \delta(z))$, and then
\begin{equation}
u_\mathrm{l}(z) \equiv u(x_\mathrm{l}, z) = -U[1 - \delta'(z)], \qquad h < z \leq H, \label{eq:u_l_a_appendix}
\end{equation}
where $\delta' \equiv \mathrm{d} \delta / \mathrm{d} z$ and we have used Eqs.\,\eqref{eq:stream_function_u} and \eqref{eq:psi_r}. Note that $u_\mathrm{l}$ and $\rho_\mathrm{l}$ may not be continuous at the interface $z=h$ according to Eqs.\,\eqref{eq:rho_l_c}, \eqref{eq:u_l_c}, \eqref{eq:rho_l_a_appendix}, and \eqref{eq:u_l_a_appendix}.

It is straightforward to examine that $\mathrm{d} \rho / \mathrm{d} \psi$ is also constant along each streamline. As a result,
\begin{equation}
\frac{\mathrm{d} \rho}{\mathrm{d} \psi}(x_\mathrm{l},z) = \frac{\mathrm{d} \rho}{\mathrm{d} \psi}(x_\mathrm{r}, z - \delta(z)), \qquad h < z \leq H.
\end{equation}

Furthermore, we have
\begin{equation}
\frac{\mathrm{d} \rho}{\mathrm{d} \psi}(x_\mathrm{r}, z - \delta(z)) = \frac{1}{U} \frac{\mathrm{d} \rho_\mathrm{a}}{\mathrm{d} z}|_{z - \delta(z)}, \qquad h < z \leq H \label{eq:rho_psi}
\end{equation}
according to Eqs.\,\eqref{eq:rho_r} and \eqref{eq:psi_r}.

Substituting Eqs.\,\eqref{eq:u_l_a_appendix}--\eqref{eq:rho_psi} into Eq.\,\eqref{eq:f_conservation}, we have
\begin{equation}
\delta''(z) = \frac{g}{\rho_0 U^2} \frac{\mathrm{d} \rho_\mathrm{a}}{\mathrm{d} z}|_{z - \delta(z)} \delta(z), \qquad h < z \leq H, \label{eq:DJL_appendix}
\end{equation}
where $\delta'' \equiv \mathrm{d}^2 \delta / \mathrm{d} z^2$. The boundary conditions are
\begin{align}
& \delta(h) = h, \\
& \delta(H) = 0.
\end{align}

 Eq.\,\eqref{eq:DJL_appendix} seems a simplified version of Long's model \cite{Long1953}, but it should be noted that  $\rho U^2$ is required to be constant in the far upstream flow in Long's model, an inapplicable assumption for a stratified ambient fluid with a uniform velocity. As shown above, this requirement is circumvented by the Boussinesq approximation in this paper.
\\
\renewcommand{\theequation}{B\arabic{equation}}
\setcounter{equation}{0}


\begin{thebibliography}{100}
\expandafter\ifx\csname
natexlab\endcsname\relax\def\natexlab#1{#1}\fi
\expandafter\ifx\csname bibnamefont\endcsname\relax
  \def\bibnamefont#1{#1}\fi
\expandafter\ifx\csname bibfnamefont\endcsname\relax
  \def\bibfnamefont#1{#1}\fi
\expandafter\ifx\csname citenamefont\endcsname\relax
  \def\citenamefont#1{#1}\fi
\expandafter\ifx\csname url\endcsname\relax
  \def\url#1{\texttt{#1}}\fi
\expandafter\ifx\csname
urlprefix\endcsname\relax\def\urlprefix{URL }\fi
\providecommand{\bibinfo}[2]{#2}
\providecommand{\eprint}[2][]{\url{#2}}

\bibitem[{\citenamefont{Benjamin et~al.}(1963)\citenamefont{Benjamin}}]{Benjamin1968}
\bibinfo{author}{\bibfnamefont{T. B.} \bibnamefont{Benjamin}},
\bibinfo{journal}{J. Fluid Mech.}
\textbf{\bibinfo{volume}{31(2)}}, \bibinfo{pages}{209}
(\bibinfo{year}{1968}).

\bibitem[{\citenamefont{Simpson et~al.}(1999)\citenamefont{Simpson}}]{Simpson1999}
\bibinfo{author}{\bibfnamefont{J. E.}~\bibnamefont{Simpson}},
\bibinfo{book}{{\it Gravity Currents: In the Environment and the Laboratory, 2nd ed.}}
(\bibinfo{year}{Cambridge University Press, 1999}).

\bibitem[{\citenamefont{Meiburg et~al.}(2010)\citenamefont{Meiburg}}]{Meiburg2010}
\bibinfo{author}{\bibfnamefont{E.} \bibnamefont{Meiburg}},
\bibnamefont{and} \bibinfo{author}{\bibfnamefont{B.}~\bibnamefont{Kneller}},
\bibinfo{journal}{Annu. Rev. Fluid Mech.}
\textbf{\bibinfo{volume}{42}}, \bibinfo{pages}{135}
(\bibinfo{year}{2010}).

\bibitem[{\citenamefont{Lowe et~al.}(2005)\citenamefont{Lowe}}]{Lowe2005}
\bibinfo{author}{\bibfnamefont{R. J.} \bibnamefont{Lowe}},
\bibinfo{author}{\bibfnamefont{J. W.} \bibnamefont{Rottman}},
\bibnamefont{and} \bibinfo{author}{\bibfnamefont{P. F.}~\bibnamefont{Linden}},
\bibinfo{journal}{J. Fluid Mech.}
\textbf{\bibinfo{volume}{537}}, \bibinfo{pages}{101}
(\bibinfo{year}{2005}).

\bibitem[{\citenamefont{Shin et~al.}(2004)\citenamefont{Shin}}]{Shin2004}
\bibinfo{author}{\bibfnamefont{J. O.} \bibnamefont{Shin}},
\bibinfo{author}{\bibfnamefont{S. B.} \bibnamefont{Dalziel}},
\bibnamefont{and} \bibinfo{author}{\bibfnamefont{P. F.}~\bibnamefont{Linden}},
\bibinfo{journal}{J. Fluid Mech.}
\textbf{\bibinfo{volume}{521}}, \bibinfo{pages}{1}
(\bibinfo{year}{2004}).

\bibitem[{\citenamefont{Maurer et~al.}(2010)\citenamefont{Maurer}}]{Maurer2010}
\bibinfo{author}{\bibfnamefont{B. D.} \bibnamefont{Maurer}},
\bibinfo{author}{\bibfnamefont{D. T.} \bibnamefont{Bolster}},
\bibnamefont{and} \bibinfo{author}{\bibfnamefont{P. F.}~\bibnamefont{Linden}},
\bibinfo{journal}{J. Fluid Mech.}
\textbf{\bibinfo{volume}{647}}, \bibinfo{pages}{53}
(\bibinfo{year}{2010}).

\bibitem[{\citenamefont{Ungarish et~al.}(2006)\citenamefont{Ungarish}}]{Ungarish2006}
\bibinfo{author}{\bibfnamefont{M.} \bibnamefont{Ungarish}},
\bibinfo{journal}{J. Fluid Mech.}
\textbf{\bibinfo{volume}{548}}, \bibinfo{pages}{49}
(\bibinfo{year}{2006}).

\bibitem[{\citenamefont{White et~al.}(2008)\citenamefont{White}}]{White2008}
\bibinfo{author}{\bibfnamefont{B. L.} \bibnamefont{White}},
\bibnamefont{and} \bibinfo{author}{\bibfnamefont{K. R.}~\bibnamefont{Helfrich}},
\bibinfo{journal}{J. Fluid Mech.}
\textbf{\bibinfo{volume}{616}}, \bibinfo{pages}{327}
(\bibinfo{year}{2008}).

\bibitem[{\citenamefont{Liu et~al.}(1996)\citenamefont{Liu}}]{Liu1996}
\bibinfo{author}{\bibfnamefont{C. H.} \bibnamefont{Liu}},
\bibnamefont{and} \bibinfo{author}{\bibfnamefont{M. W.}~\bibnamefont{Moncrieff}},
\bibinfo{journal}{J. Atmo. Sci.}
\textbf{\bibinfo{volume}{53(22)}}, \bibinfo{pages}{3303}
(\bibinfo{year}{1996}).

\bibitem[{\citenamefont{Maxworthy et~al.}(2002)\citenamefont{Maxworthy}}]{Maxworthy2002}
\bibinfo{author}{\bibfnamefont{T.} \bibnamefont{Maxworthy}},
\bibinfo{author}{\bibfnamefont{J.} \bibnamefont{Leilich}},
\bibinfo{author}{\bibfnamefont{J. E.} \bibnamefont{Simpson}},
\bibnamefont{and} \bibinfo{author}{\bibfnamefont{E. H.}~\bibnamefont{Meiburg}},
\bibinfo{journal}{J. Fluid Mech.}
\textbf{\bibinfo{volume}{453}}, \bibinfo{pages}{371}
(\bibinfo{year}{2002}).

\bibitem[{\citenamefont{Britter et~al.}(1981)\citenamefont{Britter}}]{Britter1981}
\bibinfo{author}{\bibfnamefont{R. E.} \bibnamefont{Britter}},
\bibnamefont{and} \bibinfo{author}{\bibfnamefont{J. E.}~\bibnamefont{Simpson}},
\bibinfo{journal}{J. Fluid Mech.}
\textbf{\bibinfo{volume}{112}}, \bibinfo{pages}{459}
(\bibinfo{year}{1981}).

\bibitem[{\citenamefont{Faust et~al.}(1984)\citenamefont{Faust}}]{Faust1984}
\bibinfo{author}{\bibfnamefont{K. M.} \bibnamefont{Faust}},
\bibnamefont{and} \bibinfo{author}{\bibfnamefont{E. J.}~\bibnamefont{Plate}},
\bibinfo{journal}{J. Hydraul. Res.}
\textbf{\bibinfo{volume}{22(5)}}, \bibinfo{pages}{315}
(\bibinfo{year}{1984}).

\bibitem[{\citenamefont{Borden et~al.}(2013)\citenamefont{Borden}}]{Borden2013}
\bibinfo{author}{\bibfnamefont{Z.} \bibnamefont{Borden}},
\bibnamefont{and} \bibinfo{author}{\bibfnamefont{E.}~\bibnamefont{Meiburg}},
\bibinfo{journal}{Phys. Fluids}
\textbf{\bibinfo{volume}{25}}, \bibinfo{pages}{101301}
(\bibinfo{year}{2013}).

\bibitem[{\citenamefont{Azadani et~al.}(2015)\citenamefont{Azadani}}]{Azadani2015}
\bibinfo{author}{\bibfnamefont{M. M.} \bibnamefont{Nasr-Azadani}},
\bibnamefont{and} \bibinfo{author}{\bibfnamefont{E.}~\bibnamefont{Meiburg}},
\bibinfo{journal}{J. Fluid Mech.}
\textbf{\bibinfo{volume}{778}}, \bibinfo{pages}{552}
(\bibinfo{year}{2015}).

\bibitem[{\citenamefont{Azadani et~al.}(2016)\citenamefont{Azadani}}]{Azadani2016}
\bibinfo{author}{\bibfnamefont{M. M.} \bibnamefont{Nasr-Azadani}},
\bibnamefont{and} \bibinfo{author}{\bibfnamefont{E.}~\bibnamefont{Meiburg}},
\bibinfo{journal}{Q. J. Roy. Meteor. Soc.}
\textbf{\bibinfo{volume}{142}}, \bibinfo{pages}{1359}
(\bibinfo{year}{2016}).

\bibitem[{\citenamefont{Konopliv et~al.}(2016)\citenamefont{Konopliv}}]{Konopliv2016}
\bibinfo{author}{\bibfnamefont{N. A.} \bibnamefont{Konopliv}},
\bibinfo{author}{\bibfnamefont{S. G.} \bibnamefont{Llewellyn Smith}},
\bibinfo{author}{\bibfnamefont{J. N.} \bibnamefont{Mcelwaine}},
\bibnamefont{and} \bibinfo{author}{\bibfnamefont{E.}~\bibnamefont{Meiburg}},
\bibinfo{journal}{J. Fluid Mech.}
\textbf{\bibinfo{volume}{789}}, \bibinfo{pages}{806}
(\bibinfo{year}{2016}).

\bibitem[{\citenamefont{Ungarish et~al.}(2002)\citenamefont{Ungarish}}]{Ungarish2002}
\bibinfo{author}{\bibfnamefont{M.} \bibnamefont{Ungarish}},
\bibnamefont{and} \bibinfo{author}{\bibfnamefont{H. E.}~\bibnamefont{Huppert}},
\bibinfo{journal}{J. Fluid Mech.}
\textbf{\bibinfo{volume}{458}}, \bibinfo{pages}{283}
(\bibinfo{year}{2002}).

\bibitem[{\citenamefont{Ungarish et~al.}(2005)\citenamefont{Ungarish}}]{Ungarish2005a}
\bibinfo{author}{\bibfnamefont{M.} \bibnamefont{Ungarish}},
\bibinfo{journal}{Eur. J. Mech. B/Fluids}
\textbf{\bibinfo{volume}{24}}, \bibinfo{pages}{642}
(\bibinfo{year}{2005}).

\bibitem[{\citenamefont{Ungarish et~al.}(2012)\citenamefont{Ungarish}}]{Ungarish2012}
\bibinfo{author}{\bibfnamefont{M.} \bibnamefont{Ungarish}},
\bibinfo{journal}{J. Fluid Mech.}
\textbf{\bibinfo{volume}{12}}, \bibinfo{pages}{115}
(\bibinfo{year}{2012}).

\bibitem[{\citenamefont{Goldman et~al.}(2014)\citenamefont{Goldman}}]{Goldman2014}
\bibinfo{author}{\bibfnamefont{R.} \bibnamefont{Goldman}},
\bibinfo{author}{\bibfnamefont{M.} \bibnamefont{Ungarish}},
\bibnamefont{and} \bibinfo{author}{\bibfnamefont{I.}~\bibnamefont{Yavneh}},
\bibinfo{journal}{Environ. Fluid Mech.}
\textbf{\bibinfo{volume}{14}}, \bibinfo{pages}{471}
(\bibinfo{year}{2014}).

\bibitem[{\citenamefont{Ungarish et~al.}(2005)\citenamefont{Ungarish}}]{Ungarish2005b}
\bibinfo{author}{\bibfnamefont{M.} \bibnamefont{Ungarish}},
\bibinfo{journal}{J. Fluid Mech.}
\textbf{\bibinfo{volume}{535}}, \bibinfo{pages}{287}
(\bibinfo{year}{2005}).

\bibitem[{\citenamefont{Xie et~al.}(2019)\citenamefont{Xie}}]{Xie2019}
\bibinfo{author}{\bibfnamefont{C. Y.} \bibnamefont{Xie}},
\bibinfo{author}{\bibfnamefont{J. J.} \bibnamefont{Tao}},
\bibnamefont{and} \bibinfo{author}{\bibfnamefont{L. S.}~\bibnamefont{Zhang}},
\bibinfo{journal}{Phys. Rev. E}
\textbf{\bibinfo{volume}{100}}, \bibinfo{pages}{031103(R)}
(\bibinfo{year}{2019}).

\bibitem[{\citenamefont{Zhang et~al.}(2021)\citenamefont{Zhang}}]{Zhang2021}
\bibinfo{author}{\bibfnamefont{L. S.} \bibnamefont{Zhang}},
\bibinfo{author}{\bibfnamefont{J. J.} \bibnamefont{Tao}},
\bibinfo{author}{\bibfnamefont{G. H.} \bibnamefont{Wang}},
\bibnamefont{and} \bibinfo{author}{\bibfnamefont{X. J.}~\bibnamefont{Zheng}},
\bibinfo{journal}{Acta Mech. Sinica}
\textbf{\bibinfo{volume}{37}}, \bibinfo{pages}{47}
(\bibinfo{year}{2021}).

\bibitem[{\citenamefont{Global et~al.}(2001)\citenamefont{Global}}]{Global2001}
\bibinfo{author}{\bibfnamefont{G.} \bibnamefont{Yang}},
\bibinfo{author}{\bibfnamefont{H.} \bibnamefont{Xiao}},
\bibnamefont{and} \bibinfo{author}{\bibfnamefont{W.}~\bibnamefont{Tuo}},
\bibinfo{journal}{Global Alarm: Dust and sandstorms from the world's drylands, United Nations:Bangkok, Thailand}
\bibinfo{pages}{49}
(\bibinfo{year}{2001}).

\bibitem[{\citenamefont{Liu et~al.}(2021)\citenamefont{Liu}}]{Liu2021}
\bibinfo{author}{\bibfnamefont{H. Y.} \bibnamefont{Liu}},
\bibinfo{author}{\bibfnamefont{Y. X.} \bibnamefont{Shi}},
\bibnamefont{and} \bibinfo{author}{\bibfnamefont{X. J.}~\bibnamefont{Zheng}},
\bibinfo{journal}{Atmos. Chem. Phys. Discuss.}
\bibinfo{pages}{https://doi.org/10.5194/acp-2021-889}
(\bibinfo{year}{2021}).

\bibitem[{\citenamefont{Monin et~al.}(1954)\citenamefont{Monin}}]{Monin1954}
\bibinfo{author}{\bibfnamefont{A.} \bibnamefont{Monin}},
\bibnamefont{and} \bibinfo{author}{\bibfnamefont{A.}~\bibnamefont{Obukhov}},
\bibinfo{journal}{Tr. Geofiz. Inst. Akad. Nauk. SSSR}
\textbf{\bibinfo{volume}{151}}, \bibinfo{pages}{163}
(\bibinfo{year}{1954}).

\bibitem[{\citenamefont{Kader et~al.}(1990)\citenamefont{Kader}}]{Kader1990}
\bibinfo{author}{\bibfnamefont{B. A.} \bibnamefont{Kader}},
\bibnamefont{and} \bibinfo{author}{\bibfnamefont{A. M.}~\bibnamefont{Yaglom}},
\bibinfo{journal}{J. Fluid Mech.}
\textbf{\bibinfo{volume}{212}}, \bibinfo{pages}{637}
(\bibinfo{year}{1990}).

\bibitem[{\citenamefont{Cheng et~al.}(2021)\citenamefont{Cheng}}]{Cheng2021}
\bibinfo{author}{\bibfnamefont{Y.} \bibnamefont{Cheng}},
\bibinfo{author}{\bibfnamefont{Q.} \bibnamefont{Li}},
\bibinfo{author}{\bibfnamefont{D.} \bibnamefont{Li}},
\bibnamefont{and} \bibinfo{author}{\bibfnamefont{P.}~\bibnamefont{Gentine}},
\bibinfo{journal}{Phys. Rev. Fluids}
\textbf{\bibinfo{volume}{6}}, \bibinfo{pages}{034606}
(\bibinfo{year}{2021}).

\bibitem[{\citenamefont{He et~al.}(2021)\citenamefont{He}}]{He2021}
\bibinfo{author}{\bibfnamefont{X. Z.} \bibnamefont{He}},
\bibinfo{author}{\bibfnamefont{E.} \bibnamefont{Bodenschatz}},
\bibnamefont{and} \bibinfo{author}{\bibfnamefont{G.}~\bibnamefont{Ahlers}},
\bibinfo{journal}{Theor. Appl. Mech. Lett.}
\textbf{\bibinfo{volume}{11}}, \bibinfo{pages}{100237}
(\bibinfo{year}{2021}).

\bibitem[{\citenamefont{Ahlers et~al.}(2014)\citenamefont{Ahlers}}]{Ahlers2014}
\bibinfo{author}{\bibfnamefont{G.} \bibnamefont{Ahlers}},
\bibinfo{author}{\bibfnamefont{E.} \bibnamefont{Bodenschatz}},
\bibnamefont{and} \bibinfo{author}{\bibfnamefont{X. Z.}~\bibnamefont{He}},
\bibinfo{journal}{J. Fluid Mech.}
\textbf{\bibinfo{volume}{758}}, \bibinfo{pages}{436}
(\bibinfo{year}{2014}).

\bibitem[{\citenamefont{Chevalier et~al.}(2007)\citenamefont{Chevalier}}]{Chevalier2007}
\bibinfo{author}{\bibfnamefont{M.} \bibnamefont{Chevalier}},
\bibinfo{author}{\bibfnamefont{P.} \bibnamefont{Schlatter}},
\bibinfo{author}{\bibfnamefont{A.} \bibnamefont{Lundbladh}},
\bibnamefont{and} \bibinfo{author}{\bibfnamefont{D. S.}~\bibnamefont{Henningson}},
\bibinfo{journal}{Tech. Rep. TRITA-MEK 2007:07, Royal Institute of Technology, Stockholm, Sweden}
(\bibinfo{year}{2007}).

\bibitem[{\citenamefont{Constantinescu et~al.}(2014)\citenamefont{Constantinescu}}]{Constantinescu2014}
\bibinfo{author}{\bibfnamefont{G.} \bibnamefont{Constantinescu}},
\bibinfo{journal}{Environ. Fluid Mech.}
\textbf{\bibinfo{volume}{14}}, \bibinfo{pages}{295}
(\bibinfo{year}{2014}).

\bibitem[{\citenamefont{Bonometti et~al.}(2008)\citenamefont{Bonometti}}]{Bonometti2008}
\bibinfo{author}{\bibfnamefont{T.} \bibnamefont{Bonometti}},
\bibnamefont{and} \bibinfo{author}{\bibfnamefont{S.}~\bibnamefont{Balachandar}},
\bibinfo{journal}{Theor. Comput. Fluid Dyn.}
\textbf{\bibinfo{volume}{22}}, \bibinfo{pages}{341}
(\bibinfo{year}{2008}).

\bibitem[{\citenamefont{Birman et~al.}(2007)\citenamefont{Birman}}]{Birman2007}
\bibinfo{author}{\bibfnamefont{V. K.} \bibnamefont{Birman}},
\bibinfo{author}{\bibfnamefont{E.} \bibnamefont{Meiburg}},
\bibnamefont{and} \bibinfo{author}{\bibfnamefont{M.}~\bibnamefont{Ungarish}},
\bibinfo{journal}{Phys. Fluids}
\textbf{\bibinfo{volume}{19}}, \bibinfo{pages}{086602}
(\bibinfo{year}{2007}).

\bibitem[{\citenamefont{Simpson et~al.}(1979)\citenamefont{Simpson}}]{Simpson1979}
\bibinfo{author}{\bibfnamefont{J. E.} \bibnamefont{Simpson}},
\bibnamefont{and} \bibinfo{author}{\bibfnamefont{R. E.}~\bibnamefont{Britter}},
\bibinfo{journal}{J. Fluid Mech.}
\textbf{\bibinfo{volume}{94}}, \bibinfo{pages}{477}
(\bibinfo{year}{1979}).

\bibitem[{\citenamefont{Borden et~al.}(2012)\citenamefont{Borden}}]{Borden2012}
\bibinfo{author}{\bibfnamefont{Z.} \bibnamefont{Borden}},
\bibinfo{author}{\bibfnamefont{E.} \bibnamefont{Meiburg}},
\bibnamefont{and} \bibinfo{author}{\bibfnamefont{G.}~\bibnamefont{Constantinescu}},
\bibinfo{journal}{J. Fluid Mech.}
\textbf{\bibinfo{volume}{703}}, \bibinfo{pages}{279}
(\bibinfo{year}{2012}).

\bibitem[{\citenamefont{Huppert et~al.}(1980)\citenamefont{Huppert}}]{Huppert1980}
\bibinfo{author}{\bibfnamefont{H.} \bibnamefont{Huppert}},
\bibnamefont{and} \bibinfo{author}{\bibfnamefont{J. E.}~\bibnamefont{Simpson}},
\bibinfo{journal}{J. Fluid Mech.}
\textbf{\bibinfo{volume}{99}}, \bibinfo{pages}{785}
(\bibinfo{year}{1980}).

\bibitem[{\citenamefont{Hacker et~al.}(1996)\citenamefont{Hacker}}]{Hacker1996}
\bibinfo{author}{\bibfnamefont{J.} \bibnamefont{Hacker}},
\bibinfo{author}{\bibfnamefont{P. F.} \bibnamefont{Linden}},
\bibnamefont{and} \bibinfo{author}{\bibfnamefont{S. B.}~\bibnamefont{Dalziel}},
\bibinfo{journal}{Dynam. Atmos. Oceans}
\textbf{\bibinfo{volume}{24}}, \bibinfo{pages}{183}
(\bibinfo{year}{1996}).

\bibitem[{\citenamefont{Hartel et~al.}(2000)\citenamefont{Hartel}}]{Hartel2000}
\bibinfo{author}{\bibfnamefont{C.} \bibnamefont{H\"artel}},
\bibinfo{author}{\bibfnamefont{E.} \bibnamefont{Meiburg}},
\bibnamefont{and} \bibinfo{author}{\bibfnamefont{F.}~\bibnamefont{Necker}},
\bibinfo{journal}{J. Fluid Mech.}
\textbf{\bibinfo{volume}{418}}, \bibinfo{pages}{189}
(\bibinfo{year}{2000}).

\bibitem[{\citenamefont{Hallworth et~al.}(1993)\citenamefont{Hallworth}}]{Hallworth1993}
\bibinfo{author}{\bibfnamefont{M. A.} \bibnamefont{Hallworth}},
\bibinfo{author}{\bibfnamefont{J. C.} \bibnamefont{Phillips}},
\bibinfo{author}{\bibfnamefont{H. E.} \bibnamefont{Huppert}},
\bibnamefont{and} \bibinfo{author}{\bibfnamefont{R. S. J.}~\bibnamefont{Sparks}},
\bibinfo{journal}{Nature}
\textbf{\bibinfo{volume}{362}}, \bibinfo{pages}{829}
(\bibinfo{year}{1993}).

\bibitem[{\citenamefont{Hallworth et~al.}(1996)\citenamefont{Hallworth}}]{Hallworth1996}
\bibinfo{author}{\bibfnamefont{M. A.} \bibnamefont{Hallworth}},
\bibinfo{author}{\bibfnamefont{H. E.} \bibnamefont{Huppert}},
\bibinfo{author}{\bibfnamefont{J. C.} \bibnamefont{Phillips}},
\bibnamefont{and} \bibinfo{author}{\bibfnamefont{R. S. J.}~\bibnamefont{Sparks}},
\bibinfo{journal}{J. Fluid Mech.}
\textbf{\bibinfo{volume}{308}}, \bibinfo{pages}{289}
(\bibinfo{year}{1996}).

\bibitem[{\citenamefont{Li et~al.}(2021)\citenamefont{Li}}]{Li2021}
\bibinfo{author}{\bibfnamefont{X. B.} \bibnamefont{Li}},
\bibinfo{author}{\bibfnamefont{Y. X.} \bibnamefont{Huang}},
\bibinfo{author}{\bibfnamefont{G. H.} \bibnamefont{Wang}},
\bibnamefont{and} \bibinfo{author}{\bibfnamefont{X. J.}~\bibnamefont{Zheng}},
\bibinfo{journal}{Earth Syst. Sci. Data}
\textbf{\bibinfo{volume}{13}}, \bibinfo{pages}{5819}
(\bibinfo{year}{2021}).

\bibitem[{\citenamefont{Wang et~al.}(2017)\citenamefont{Wang}}]{Wang2017}
\bibinfo{author}{\bibfnamefont{G. H.} \bibnamefont{Wang}},
\bibinfo{author}{\bibfnamefont{X. J.} \bibnamefont{Zheng}},
\bibnamefont{and} \bibinfo{author}{\bibfnamefont{J. J.}~\bibnamefont{Tao}},
\bibinfo{journal}{Phys. Fluids}
\textbf{\bibinfo{volume}{29}}, \bibinfo{pages}{061701}
(\bibinfo{year}{2017}).

\bibitem[{\citenamefont{Srivastava et~al.}(2011)\citenamefont{Srivastava}}]{Srivastava2011}
\bibinfo{author}{\bibfnamefont{A. K.} \bibnamefont{Srivastava}},
\bibinfo{author}{\bibfnamefont{P.} \bibnamefont{Pant}},
\bibinfo{author}{\bibfnamefont{P.} \bibnamefont{Hedge}},
\bibinfo{author}{\bibfnamefont{S.} \bibnamefont{Singh}},
\bibinfo{author}{\bibfnamefont{U. C.} \bibnamefont{Dumka}},
\bibinfo{author}{\bibfnamefont{M.} \bibnamefont{Naja}},
\bibinfo{author}{\bibfnamefont{N.} \bibnamefont{Singh}},
\bibnamefont{and} \bibinfo{author}{\bibfnamefont{Y.}~\bibnamefont{Bhavanikumar}},
\bibinfo{journal}{Int. J. Remote Sens.}
\textbf{\bibinfo{volume}{32}}, \bibinfo{pages}{7827}
(\bibinfo{year}{2011}).

\bibitem[{\citenamefont{Long et~al.}(1953)\citenamefont{Long}}]{Long1953}
\bibinfo{author}{\bibfnamefont{R. R.} \bibnamefont{Long}},
\bibinfo{journal}{Tellus}
\textbf{\bibinfo{volume}{5(1)}}, \bibinfo{pages}{42}
(\bibinfo{year}{1953}).

\end{thebibliography}

\end{document}